\documentclass[10pt]{aastex}
\usepackage{emulateapj5}
\usepackage{apjfonts}
\usepackage{psfig}

\newcommand{\kms}       {\mbox{km s$^{-1}$}}%
\newcommand{\kmsMpc}	{\mbox{km s$^{-1}$ Mpc$^{-1}$}}%
\newcommand{\dbr}	{\mbox{$\Delta(B\!-\!R)$}}

\shortauthors{Kannappan, Jansen, \& Barton}
\shorttitle{Forming Young Bulges}

\slugcomment{Accepted by AJ 2003 December 16}

\begin{document}

\title {Forming Young Bulges within Existing Disks: \\
Statistical Evidence for External Drivers}
\author{Sheila J. Kannappan\altaffilmark{1},
	Rolf A. Jansen\altaffilmark{2},
	and Elizabeth J. Barton\altaffilmark{3}}
\altaffiltext{1}{Harlan Smith Fellow, McDonald Observatory, The
	University of Texas at Austin, 1 University Station C1402,
	Austin, TX 78712-0259; sheila@astro.as.utexas.edu}
\altaffiltext{2}{Department of Physics and Astronomy, Arizona State
	University, Box 871504, Tempe, AZ 85287-1504; Rolf.Jansen@asu.edu}
\altaffiltext{3}{Hubble Fellow, Steward Observatory, University of
	Arizona, 933 N. Cherry Ave., Tucson, AZ 85721; ebarton@as.arizona.edu}

\begin{abstract}

Contrary to traditional models of galaxy formation, recent observations
suggest that some bulges form within preexisting disk galaxies.  Such
late-epoch bulge formation within disks seems to be linked to disk gas
inflow and central star formation, caused by either internal secular
processes or galaxy mergers and interactions.  We identify a population of
galaxies likely to be experiencing active bulge growth within disks, using
the criterion that the color within the half-light radius is bluer than the
outer disk color.  Such blue-centered galaxies make up $>$10\% of
star-forming disk galaxies within the Nearby Field Galaxy Survey, a broad
survey designed to represent the natural diversity of the low-$z$ galaxy
population over a wide range of luminosities and environments.
Blue-centered galaxies correlate at 99\% confidence with morphological
peculiarities suggestive of minor mergers and interactions.  From this and
other evidence, we argue that external drivers rather than internal secular
processes probably account for the majority of blue-centered galaxies.  We
go on to discuss quantitative plausibility arguments indicating that
blue-centered evolutionary phases may represent an important mode of bulge
growth for most disk galaxies, leading to significant changes in
bulge-to-disk ratio without destroying disks.  If this view is correct,
bulge growth within disks may be a natural consequence of the repeated
galaxy mergers and interactions inherent in hierarchical galaxy formation.

\end{abstract}


\section{Introduction}
\label{sc:intro}

Traditional wisdom holds that galaxy bulges are old, red, and spheroidal.
In this view, bulges are essentially tiny elliptical galaxies that formed
early on, via either primordial collapse or hierarchical mergers
\citep[e.g.][]{eggen.lynden-bell.ea:evidence,kauffmann:age}.  The fact that
we call these small spheroids ``bulges,'' according to this picture, simply
reflects the fact that they later accreted disks.

However, bulges can also be young, blue, and disky.  Some bulges display
offsets from the Faber-Jackson relation \citep{faber.jackson:velocity} that
are indicative of youthful mass-to-light ratios or disky kinematics
\citep{kormendy.illingworth:l,kormendy:kinematics}.  Studies of bulge
colors suggest younger ages and/or lower metallicities for late-type galaxy
bulges compared to early-type galaxy bulges
\citep{peletier.balcells.ea:galactic,carollo.stiavelli.ea:hubble}, and for
bulges compared to elliptical galaxies \citep{ellis.abraham.ea:relative}.
Bulge colors and inner disk colors correlate closely, hinting at connected
formation histories \citep{peletier.balcells:ages}.  Late-type galaxy
bulges tend to have exponential rather than r$^{1/4}$-law profiles
\citep[e.g.,][]{andredakis.peletier.ea:shape}, with bulge scale lengths
possibly linked to outer disk scale lengths \citep[][ and references
therein]{courteau..ea:evidence,macarthur.courteau.ea:structure}.
\citet{prugniel.maubon.ea:formation} find that the Mg$_{2}$ index for
Sa--Sc bulges correlates better with the outer disk rotation velocity than
with the bulge velocity dispersion.

This body of evidence implies that many bulges probably formed or grew
substantially at late epochs, evolving both within and together with
preexisting disks ({\em ``in situ''} bulge formation).  Two possible
scenarios for {\em in situ} bulge formation exist.  One possibility is
internal secular evolution driven by bar instabilities or other
non-axisymmetric distortions, which can cause disk gas inflow, central star
formation that enhances the bulge-to-disk ratio, and vertical resonances
that diffuse disk stars into the bulge
\citep[e.g.,][]{pfenniger.norman:dissipation, friedli.benz:secular}.
Alternatively, the same phenomena --- non-axisymmetric distortions, disk
gas inflow, and central star formation --- can be triggered by neighbor
interactions or minor mergers
\citep[e.g.,][]{mihos.hernquist:triggering,steinmetz.navarro:hierarchical}.
Even a 10$\times$ lower-mass companion can trigger disk gas inflow far
disproportionate to the perturber's mass
\citep{hernquist.mihos:excitation}.  Accretion of small companions can also
contribute to {\em in situ} bulge growth directly
\citep[e.g.,][]{walker.mihos.ea:quantifying,aguerri.balcells.ea:growth}.
In studies of individual galaxies and non-statistical samples, internally
and externally driven scenarios for {\em in situ} bulge growth are
virtually completely degenerate.  One can rarely rule out faint companions
over a wide field,
and evidence for minor mergers is even more difficult to detect and
interpret.

In this paper, we focus on a class of galaxies likely to be currently
building bulges via {\em in situ} mechanisms, and we attempt to resolve the
internal/external trigger degeneracy for these galaxies by statistical
means.  We identify likely examples of {\em in situ} bulge growth by a
purely operational criterion adapted from \citet{jansen.franx.ea:surface}:
we look for galaxies in which the color within the half-light radius is
bluer than the outer disk color.  In such ``blue-centered'' galaxies, newly
formed stars dominate the central colors and may plausibly be changing the
concentration index and bulge-to-disk ratio, through both temporary decreases
in central stellar mass-to-light ratio and permanent build-up of stellar
mass at the center
\citep[][]{schweizer:interactions,wyse.gilmore.ea:galactic,barton-gillespie.geller.ea:tidally}.
The blue-center criterion allows us to examine strongly evolving systems
that may contain bars, merger debris, or other unrelaxed structures that
preclude reliable bulge-to-disk decomposition.  We show that the spatial
scales of star formation in these objects are consistent with bulge growth.

To determine the frequency and physical drivers of blue-centered
galaxies in the general galaxy population, we examine their
distribution within the Nearby Field Galaxy Survey
\citep[NFGS,][]{jansen.fabricant.ea:spectrophotometry,jansen.franx.ea:surface,kannappan:kinematic},
a statistically representative sample of low-redshift galaxies
spanning all Hubble types over a wide range of luminosities and
environments.  Blue-centered galaxies make up $>$10\% of star-forming
disk galaxies brighter than M$_{\rm B}=-17.5$ in the NFGS, and they
occur with increasing frequency at lower luminosities
\citep[][]{tully.verheijen.ea:ursa,jansen.franx.ea:surface}.  Based on
independent, uniform morphological peculiarity classifications for the
survey, we will argue that external triggers probably account for the
majority of blue-centered galaxies in the NFGS.  Uniform but shallow
companion data support the same result, with some interpretive
ambiguity.  This analysis extends previous work linking galaxy
encounters to starbursts and central star formation
\citep[e.g.,][]{larson.tinsley:star,kennicutt.keel:induced,keel.kennicutt.ea:effects,schweizer:interactions,barton.geller.ea:tidally,barton-gillespie.geller.ea:tidally}
in that we assess how often central star formation can be traced to
galaxy encounters as opposed to other processes at work in the general
galaxy population.  Of course, we consider only a special class of
central star formation, intense and widespread enough to cause a blue
center.  The blue-center criterion is not sensitive to small-scale
nuclear star formation, star formation on top of very red underlying
populations, or central star formation surrounded by comparable
amounts of disk star formation.  Nonetheless, both the abundance of
blue-centered galaxies and the intensity and extent of their central
star formation suggest that understanding these galaxies may be
important to understanding the phenomenon of {\em in situ} bulge
growth.

The remainder of this paper is organized as follows.  In
\S~\ref{sc:samples} we briefly describe the Nearby Field Galaxy Survey as
well as a second sample, the Close Pairs Survey
\citep{barton.geller.ea:tidally}, which we use for further analysis of
blue-centered galaxies with companions.  In \S~\ref{sc:bcid} we define the
outer-minus-inner color difference \dbr, discuss \dbr\ trends with
luminosity and morphology, and summarize the properties of blue-centered
galaxies within the NFGS.  In \S~\ref{sc:drivers} we present statistical
correlations between blue-centered galaxies (as well as \dbr) and factors
that may indicate external disturbance, i.e., morphological peculiarities
and close companions.  In \S~\ref{sc:plausibility} we present a series of
quantitative plausibility arguments to support the suggestion that
blue-centered phases represent an important mode of {\em in situ} bulge
growth.  Finally, in \S~\ref{sc:openq} we discuss open questions raised by
these results regarding the relationship of blue-centered galaxies to the
phenomenon of disky bulges and to the paradigm of hierarchical galaxy
formation.  We summarize our results in \S~\ref{sc:concl}.

\section{Survey Data}
\label{sc:samples}

The Nearby Field Galaxy Survey is ideal for this analysis because it was
designed to provide a statistically representative sample of the general
galaxy population, without explicit selection bias in color, morphology, or
any other galaxy property \citep[][ hereafter
J00]{jansen.franx.ea:surface}.  Imaging, spectrophotometry, and kinematic
data for the NFGS come from J00,
\citet{jansen.fabricant.ea:spectrophotometry}, and
\citet{kannappan:kinematic}, respectively. The 196 galaxies in the NFGS
were drawn from the CfA~1 redshift survey \citep{huchra.davis.ea:survey} in
numbers roughly proportional to the local galaxy luminosity function and
span the full range of Hubble types and environments (with the exclusion of
the Virgo Cluster to avoid overrepresentation of cluster galaxies).
However, a key point for the analysis in this paper is the fact that the
properties of the NFGS do not reflect those of a flux-limited sample. More
luminous galaxies were selected at progressively larger distances (as
described in detail in J00), such that all apparent magnitudes $m_Z$ fall
in a very narrow range 14--14.5, set by the CfA~1 magnitude limit (where
$m_Z$, the Zwicky magnitude, is similar to a B-band magnitude).  In turn,
the CfA~1 survey is a subset of the Updated Zwicky Catalog
\citep[UZC,][]{falco.kurtz.ea:updated}, which is an edited and updated
version of the CfA~2 redshift survey, with magnitude limit $m_Z=15.5$.
Therefore the UZC allows us to perform systematic searches for companions
to NFGS galaxies down to $\sim$1~mag fainter than the primary.  This type
of search avoids distance-dependent sampling biases: rather than searching
for companions down to a specified absolute magnitude, we search for
companions down to $\sim$1 mag fainter than the primary, regardless of
primary luminosity or redshift.  Physically, such a search is sensible, as
the detection threshold reflects an approximately constant interaction mass
ratio, regardless of absolute mass scale.

Unfortunately, as we will see in \S~\ref{sc:driversa}, many of the
companions relevant to blue-centered galaxies may be too faint, or too
closely blended, to be separately catalogued in the UZC.  To detect
disturbances from small companions or mergers already in progress, we use
the uniform set of morphological peculiarity flags defined for the NFGS by
\citet{kannappan.fabricant.ea:physical}.  These flags do not imply a type
``Pec'' classification.  To avoid confusion on this point, we group both
type ``Im'' and type ``Pec'' galaxies under type ``Irr'' in this
paper. This redefinition is incidental, because except for a brief look at
the full NFGS in \S~\ref{sc:bcid}, we will focus most of our attention on a
sample that excludes nearly all type Irr galaxies via a magnitude cut at
M$_{\rm B}=-17.5$.

We also draw on a second survey, the 191-galaxy Close Pairs Survey of
\citet{barton.geller.ea:tidally}, to supplement our analysis of
blue-centered galaxies known to have close companions.  The Close Pairs
Survey was selected from the CfA~2 redshift survey
\citep{geller.huchra:mapping} as a statistical sample of galaxy pairs with
line-of-sight velocity separation $\Delta V < 1000$ \kms\ and projected
spatial separation $\Delta X \la 100$ kpc \citep[][ the $\Delta X$ limit
differs from the original reference because we have recomputed all
distances as discussed below]{barton.geller.ea:tidally}.  Unlike the NFGS,
the Close Pairs Survey reflects the inherent luminosity distribution of its
magnitude-limited parent survey.  Morphological types for the Close Pairs
Survey are based on independent estimates by each of us, using the NFGS as
a common calibration sample.\footnote{We have reclassified one NFGS galaxy,
A11332+3536, from S0 to SB0/a.}

All magnitudes and distances in this paper are computed using H$_0=75$
\kmsMpc, with a correction for Virgocentric infall based on the model of
\citet[][ see J00]{kraan-korteweg.sandage.ea:effect}.  We use Galactic
extinction corrections derived from \citet{schlegel.finkbeiner.ea:maps} for
both surveys, and we quote magnitudes without internal extinction
corrections except where otherwise indicated.  We have checked that adding
internal extinction corrections does not influence our results.

\section{Characterizing Blue-Centered Galaxies}
\label{sc:bcid}

Fig.~1 shows the images and surface brightness profiles for all
blue-centered galaxies brighter than M$_{\rm B}=-17.5$ in the NFGS.  We
identify these galaxies by the criterion \dbr\ $>$ 0, where the color
difference \dbr\ equals the outer disk color (from the half-light radius
$r_e$ to the 75\%-light radius) minus the central color (averaged within
$r_e$).  Based on simulated measurements of \dbr\ for model galaxies, this
definition of \dbr\ recovers the disk-minus-bulge color difference with
less sensitivity to variations in bulge-to-disk ratio than the definition
of \dbr\ used by J00.  We exclude galaxies with formal errors in \dbr\ $>$
0.15 mag as well as galaxies with Seyfert 1 or BL Lac nuclei, for which we
cannot measure the color of the underlying stellar population (Fig.~2).
Typical errors in \dbr\ are $\sim$0.05 mag.

Fig.~2 displays the luminosity and morphology distribution of blue-centered
galaxies within the NFGS as a whole.  Blue-centered galaxies become more
common at lower luminosities \citep[][ J00]{tully.verheijen.ea:ursa}.  The
NFGS shows a natural deficit of intermediate types at low luminosities,
reflecting a genuine trend in the general galaxy population, but within
this trend, blue-centered galaxies show no preferred morphology.  A
two-sided K-S test on the morphology distributions for blue-centered and
non-blue-centered galaxies in the range $-17.5<{\rm M}_{\rm B}<-20$ yields
58\% probability that they were drawn from the same parent distribution,
confirming that there is no correlation.

For the remainder of this paper, we restrict our analysis to galaxies
brighter than M$_{\rm B}=-17.5$.  This limit excludes dwarf galaxies for
which measurements of \dbr\ might be dominated by a few stochastically
distributed star formation regions (J00).  Relaxing this limit to include
fainter galaxies yields stronger statistical results but complicates their
interpretation.

The low luminosities of blue-centered galaxies reflect a more general
correlation between the continuous parameter \dbr\ and luminosity, followed
by the entire population of star-forming\footnote{We identify galaxies as
star forming if the high-resolution spectra from the NFGS kinematic
database \citep{kannappan:kinematic} show emission lines beyond the
nucleus.  Due to lower resolution or missing data, not all of these
galaxies are listed as having emission in the NFGS spectrophotometric
database \citep{jansen.fabricant.ea:spectrophotometry}.}  disk galaxies
(Fig.~3a, Spearman rank test probability of no correlation
$1\times10^{-4}$).  While we have chosen the \dbr\ $>$ 0 criterion as a
conservative way to identify galaxies whose central concentrations are
growing, some of the almost-blue-centered outliers at the bright end of the
luminosity--\dbr\ relation may also be experiencing preferential central
growth, diluted in color by the redness of a large preexisting bulge.  In
fact, blue-centered galaxies lie at one extreme of an approximately
Gaussian distribution of \dbr\ values for star-forming disk galaxies
(Fig.~3b).  Clearly the processes occurring in blue-centered galaxies may
also affect non-blue-centered galaxies to various degrees.  We return to
this point in \S~\ref{sc:driversa} and~\ref{sc:plausimp}.

As seen in Figs.~2--3, nearly all blue-centered galaxies in the NFGS
are star-forming disk galaxies.  This fact is unsurprising, as nearly
all NFGS galaxies fainter than M$_{\rm B}=-20$ are star-forming disk
galaxies.  Most blue-centered galaxies are also starburst galaxies:
the eleven in our NFGS sample have median central and global H$\alpha$
equivalent widths of 37 and 29 \AA, respectively.\footnote{Central
equivalent widths were extracted from a fixed $3\arcsec\times7\arcsec$
aperture, while global equivalent widths were obtained by scanning the
slit across the galaxy \citep{jansen.fabricant.ea:spectrophotometry}.}
Individual central equivalent widths reach as high as $\sim$130\AA.
Only one galaxy, NGC~5541, has a central EW(H$\alpha$) below
$\sim$15\AA\ in emission; this galaxy is also exceptional among
blue-centered galaxies for its high luminosity, and its blue-centered
status may partly reflect blending with a smaller companion (see
Appendix~A).  Because we have taken an objective, operational approach
to identifying blue-centered galaxies, we choose not to reject
NGC~5541 from our sample.  However, we have verified that all of our
results hold with comparable statistical significance if this galaxy
is excluded.

\section{What Drives the Star Formation?}
\label{sc:drivers}

\subsection{Local Environment}
\label{sc:driversa}

In this section we consider how blue-centered galaxies, as well as the
continuous parameter \dbr, correlate with possible evidence for
interactions and mergers.  We expect that interactions {\em can} produce
blue-centered galaxies, as numerous studies have demonstrated that galaxy
encounters can enhance central star formation
\citep[e.g.,][]{keel.kennicutt.ea:effects,barton.geller.ea:tidally}.
Confirming this expectation, Fig.~4 shows that blue-centered galaxies in
the Close Pairs Survey tend to have closer companions than do their
non-blue-centered counterparts.  Also, blue-centered galaxies are roughly
twice as common in the Close Pairs Survey as in the NFGS ($\sim$20\% vs.\
$\sim$10\% of star-forming disk galaxies brighter than M$_{\rm B}=-17.5$),
despite the top-heavy luminosity distribution of the Close Pairs Survey
(Fig.~5).  However, these results do not address the question of whether
galaxy interactions are the {\em primary} cause of blue-centered galaxies
in the general galaxy population or just one possible cause.

To understand the source of blue-centered galaxies in the general galaxy
population, we turn to the NFGS.  Our analysis considers two types of
evidence for interactions and mergers: morphological peculiarities and
close companions.  We use the uniform peculiarity flags that were defined
for the NFGS by \citet{kannappan.fabricant.ea:physical} in the context of
identifying galaxies that might be considered disturbed from the point of
view of analyzing the Tully-Fisher relation \citep{tully.fisher:new}.
These flags may indicate multiple nuclei, likely interacting companions,
tidal tails or debris, or asymmetries not readily attributable to late-type
morphology.  Such features often provide the only direct evidence of a
merger origin for starburst activity
\citep[e.g.,][]{schweizer.seitzer:correlations}.  Although inevitably
subjective, the NFGS peculiarity flags combine the independent judgement of
two classifiers as well as notes in the literature, and reflect a uniform
peculiarity threshold.  Unfortunately, compared to the human eye,
quantitative measures such as the photometric asymmetry index of
\citet{abraham.tanvir.ea:galaxy} do not perform well in detecting minor
disturbances associated with small companion
interactions.\footnote{Kinematic asymmetries may be a somewhat more
powerful tool
\citep[e.g.,][]{kannappan:kinematic,kannappan.fabricant.ea:physical,kannappan.barton:tools},
but separating interaction-induced kinematic asymmetries from other effects
requires highly inclined galaxies with spatially well-sampled rotation
curves.} All of the blue-centered galaxies in Fig.~1, including NGC~5541,
have unexceptional photometric asymmetries \citep[as measured by][
following the methods of Abraham et al.]{jansen:nearby}.  In fact, within
the NFGS, only a handful of extreme major merger remnants are outliers in
plots of photometric asymmetry vs.\ color or luminosity, consistent with
the results of \citet{conselice:relationship}, who finds unusual
asymmetries only for severely disturbed Antennae-like systems.

We identify companions within 300 kpc and 300 \kms\ of each NFGS
galaxy using the UZC (\S~\ref{sc:samples}).  The value 300 \kms\ was chosen
because most blue-centered galaxies in the Close Pairs Survey have
companion separations $\la$300 \kms\ (Fig.~4).  Assuming a starburst
timescale of $\sim$1 Gyr, we infer that a companion as distant as $\sim$300
kpc away might have caused the observed starburst in a blue-centered
galaxy.  The UZC includes galaxies as much as 1 mag fainter than the CfA~1
redshift survey, the parent survey of the NFGS.  However, we find that
several blue-centered galaxies have possible companions that are too close
or too faint to be catalogued separately in the UZC; these companions do
not enter in our statistical analysis of companions, but we discuss them
further below.  Our results suggest that a much deeper (future) catalog
will be necessary to detect the majority of relevant companions and perform
decisive statistical tests.

Blue-centered galaxies with morphological peculiarities or UZC companions
are labeled {\em p} and {\em u} in Fig.~1.  From simple binomial
statistics, peculiarities and UZC companions are overabundant among
blue-centered galaxies at 99\% and 95\% confidence, respectively, compared
to star-forming disk galaxies in general (Table~1).  Combining the two
types of evidence, 9 of 11 blue-centered galaxies show either a peculiarity
or a UZC companion or both, indicating a correlation at 99.9\% confidence.

Similar but weaker correlations emerge if we treat \dbr\ as a continuous
parameter.  Fig.~6 plots the \dbr\ distributions for star-forming disk
galaxies (a) with and without strong peculiarities and (b) with and without
UZC companions.  Kolmogorov-Smirnov tests indicate a 1.6\% probability that
peculiar and non-peculiar galaxies have the same parent \dbr\ distribution,
and a 12\% probability that galaxies with and without UZC companions have
the same parent \dbr\ distribution.

The peculiarity--\dbr\ correlation is independent of both the UZC
companion--\dbr\ correlation and the luminosity--\dbr\ correlation
discussed in \S~\ref{sc:bcid}.  If anything, Fig.~7a shows a slight bias
{\em against} identifying strong peculiarities in low-luminosity galaxies,
presumably because many low-luminosity galaxies have late-type morphologies
that make peculiarities less obvious.  Furthermore, the peculiarity
correlation actually strengthens if we define a luminosity-corrected
parameter, \dbr$_{corr}$, equal to the residuals from the relation in
Fig.~3a.  Physically, the introduction of \dbr$_{corr}$ is motivated by the
possibility that brighter galaxies may more often have large preexisting red
bulges or bars that dilute the color difference \dbr\ even in the presence of
preferential central star formation.  Peculiar and non-peculiar galaxies
differ in their \dbr$_{corr}$ distributions at 99.5\% confidence in a K-S
test (Fig.~8).

In contrast, the presence or absence of a UZC companion does not clearly
affect the distribution of \dbr$_{corr}$.  In fact, the weak correlation
between UZC companions and the uncorrected color difference \dbr\ may not
be entirely independent of the luminosity--\dbr\ correlation, because UZC
companions are more common for lower luminosity galaxies (Fig.~7b; K-S test
confidence = 98\%).  This result is not unique to the NFGS; we have
verified it for the entire UZC.\footnote{If we select as primaries all UZC
galaxies brighter than B=14.5 (i.e. 1 mag above the survey limit), then
search for their companions in the full UZC (all the way down to 15.5), we
find a K-S test probability of $6.6\times10^{-29}$ that primaries with and
without companions share the same underlying luminosity distribution. The
subsample with companions clearly includes a higher proportion of
lower-luminosity primaries, just as for the NFGS in Fig. 7.  Furthermore,
when we perform a second test considering only fainter companions between
14.5 and 15.5, the K-S result disappears, indicating that the excess
companions detected for lower-luminosity primaries in the first test are
excess {\em brighter} companions.  These results are consistent with the
argument that the luminosity--companion correlation is caused by the
bright-end cutoff of the luminosity function: as the primary luminosity
increases, there is a decrease in the physically allowed range of companion
luminosities within which companions can be brighter than the primary.}  As
discussed in \S~\ref{sc:samples}, our companion search is not subject to
distance-dependent sampling biases.  Physically, the UZC
companion--luminosity correlation seems to reflect the fact that the galaxy
luminosity function cuts off at high luminosities: brighter neighbors are
more common for faint primaries than for bright primaries, because
faint primaries have a wider range of possible neighbor luminosities
within which neighbors {\em can} be brighter than the primary.  These
brighter neighbors may well trigger some blue-centered starbursts.
However, we suspect that the luminosity--\dbr\ correlation is more
fundamental than the UZC companion--\dbr\ correlation, for two reasons: (1)
the former is statistically stronger than the latter (although the discrete
correlations involving blue-centered galaxies instead of \dbr\ are of
comparable, weaker strength); and (2) the Close Pairs Survey, a sample in
which all galaxies have companions, shows a clear luminosity--\dbr\
correlation (as suggested by Fig.~5).

We conclude that the link between morphological peculiarities and
blue-centered galaxies is robust, while the link between UZC companions and
blue-centered galaxies is not.  Most likely, the UZC misses the majority of
relevant companions, because they are too faint, too close, or already
merging.  Four of the eleven blue-centered galaxies in our sample show
evidence of possible close or merging companions that are not listed in the
UZC (Appendix~A: A10504+0454, NGC~3846, NGC~5541, and NGC~5875A).
Morphological peculiarities may sometimes indicate the presence of such
companions, but not always.  Of the four galaxies just mentioned, only two
are flagged as clearly peculiar based on NFGS images.  For the other two,
evidence of a companion relies on new, much higher-resolution imaging and
infrared color information.  Adding to this, we find that two other
blue-centered galaxies whose NFGS images do not reveal peculiarites also
show evidence of recent merging, based on kinematic data or again, new much
higher-resolution imaging (Appendix~A: A12001+6439 and A11332+3536).  While
these latter two galaxies actually do have companions in the UZC, they may
well be minor merger cases where the UZC companions are incidental.

Overall, we find that every one of our blue-centered galaxies shows some
evidence of interactions and mergers, often suggestive of small companion
accretion.  However, a deep high-resolution survey capable of uniformly
detecting the more subtle signs of such encounters does not yet exist, so
for the present our statistical claims cannot go beyond what we have shown
using UZC companion data and the uniform peculiarity flags derived from the
original NFGS images.

\subsection{Global Environment}

On larger scales, blue-centered galaxies in the NFGS and the Close Pairs
Survey do not seem to favor any particular environment (Fig.~9).  A K-S
test shows that the environmental density distributions of blue-centered
and non-blue-centered galaxies are not statistically different.  The
majority of blue-centered galaxies we find, like the majority of galaxies
in our surveys, are in moderate-to-low density environments.  Low-density
{\em global} environments may provide ideal conditions for starbursts to be
triggered by {\em local} interactions: lower speed encounters, more cold
gas available to form stars, and later epochs of hierarchical assembly
\citep[as found by][]{grogin.geller:imaging*1}.  We note that the S0 and
S0/a blue-centered galaxies in our surveys do not reside in dense
environments and probably do not reflect cluster processes such as
harassment or ram-pressure stripping.
Our surveys include relatively few galaxies in the high-density
environments where such processes might play a larger role.

\newpage

\subsection{Bars}

Blue-centered galaxies show only a slight, statistically insignificant
excess of bars compared to the general population of star-forming disk
galaxies (Table~1).  Likewise, the presence of bars does not correlate with
\dbr\ or with central or global EW(H$\alpha$), except in the broad sense
that bars occur predominantly in star-forming disk galaxies.  Moreover, we
detect no correlation between bars and either local or global environment
within the star-forming disk galaxy sample.

These results appear to challenge both the secular and the
interaction-driven scenarios for {\em in situ} bulge growth, because
both scenarios predict bar formation that drives disk gas inflow.
However, bar statistics should be interpreted with caution, given the
possibly heterogeneous origins of bars as well as their long lifetimes
\citep[e.g.,][]{noguchi:barred,sellwood:most}.  In addition,
small-scale bars may be difficult to detect at the
$1\farcs5$--2$\arcsec$ spatial resolution of the images in Fig.~1, and
distortions from interactions and minor mergers may make bars
difficult to identify.

We also note that although previous studies have shown that bars enhance
{\em nuclear} star formation \citep[see][ and references
therein]{kennicutt:star}, such star formation may have insufficient
intensity or spatial extent to influence blue-centered galaxy statistics.
It remains an open question whether enhanced central star formation in the
general galaxy population is frequently or predominantly associated with
bars.

\section{Blue-Centered Galaxies as an Important Mode of In Situ Bulge Growth}
\label{sc:plausibility}

We have demonstrated that in a statistically representative galaxy sample,
most if not all blue-centered galaxies show evidence of interactions and
mergers.  Here we offer plausibility arguments for the further claim that
blue-centered (or almost-blue-centered) phases represent an important mode
of {\em in situ} bulge growth.

\subsection{Linking Blue Centers to Young Bulges}
\label{sc:plausbulge}

Linking central star formation to bulge formation requires a working
definition of the word ``bulge.''  We know that many bulges defy the
stereotype of smooth, red, $r^{1/4}$-law spheroids.  Improved
resolution and/or subtraction of the smooth underlying stellar
distribution often reveal complex internal structures such as bars,
spiral arms, and knots \citep{kormendy:kinematics,carollo:centers}.
Sersic $r^{1/n}$-law fits to bulge surface brightness profiles yield
$n$ values ranging from the classical $n=4$ (or even larger) down to
$n<1$ \citep{andredakis.peletier.ea:shape}, where $n=1$ describes an
exponential profile.  In some cases, the basic distinction between
bulges and disks may turn out to be an artifact of traditional
analysis techniques: \citet{b-oker.stanek.ea:searching} report that
late-type galaxies observed with HST resolution are equally well
fitted with a single Sersic profile as with two exponentials.
\citet{wyse.gilmore.ea:galactic} offer the practical definition that a
bulge is ``any light that is in excess of an inward extrapolation of a
constant scale-length exponential disk,'' but even this definition
falls short when the surface brightness profile of the outer disk is
not well described by a single exponential (e.g., when the profile is
disturbed or contains a second disk or shelf).

In light of these arguments, we suggest that any central excess of
stars on a spatial scale traditionally associated with bulges may be
plausibly described as a bulge or proto-bulge structure.  Even bars,
when they have the right spatial scale, may be identified with bulges,
as buckled or destroyed bars are expected to evolve into bulge
components \citep[e.g.,][]{friedli.benz:secular}.  Therefore a good
consistency check on whether central star formation implies bulge
formation would be to verify that the spatial scale of the star
formation corresponds to a typical bulge size scale.  Fully formed
bulges loosely scale according to $r_e^{bulge}\sim0.2r_e^{disk}$ and
$r_e^{bulge}\sim0.1r_e^{disk}$ for early and late type spiral
galaxies, respectively, with values up to
$r_e^{bulge}\sim0.4r_e^{disk}$ observed
\citep{moriondo.giovanardi.ea:near-infrared,
graham:investigation,macarthur.courteau.ea:structure}.  Also, since
fully formed bulges and inner disks show evidence of closely related
star formation histories \citep[e.g.,][]{peletier.balcells:ages}, the
starbursts in blue-centered galaxies may extend to both bulges and
inner disks.  We therefore adopt the upper envelope of the
observations, $r_e^{bulge} = 0.4r_e^{disk}$, as a plausible estimate
of the expected scale of a bulge-building starburst.

The shaded regions of the color profiles in Fig.~1 show the regions
within $r_e^{bulge} = 0.4r_e^{disk}$ for the 11 blue-centered galaxies
in our analysis.  For most galaxies, we measure $r_e^{disk}$ from an
exponential fit to the outer-disk surface brightness profile.  For a
few galaxies with ill-defined outer disks (those with a ``?'' in
Fig.~1), we compute $r_e^{bulge}$ using $r_e^{tot}$ rather than
$r_e^{disk}$.  When this substitution is necessary, the computed
$r_e^{bulge}$ is a lower limit.

Inspection of the color profiles in Fig.~1 reveals that in most cases,
the shaded regions and the regions of excess blue light correspond
reasonably well.  (NGC~5541 is a strong exception, possibly for the
reasons discussed in \S~\ref{sc:bcid}.)  Quantifying this
correspondence is tricky however.  To estimate the spatial scale of
the excess blue light very approximately, we add the $B-R$ color of
the outer disk to the R-band profile and subtract the result from the
B-band profile.  We then compute the half-light radius $r_e^{blue}$ of
the excess blue light.  Blue-centered galaxies have a median
$r_e^{blue}$ of $\sim$0.5$r_e^{disk}$ (where again we substitute
$r_e^{tot}$ for $r_e^{disk}$ for galaxies with ill-defined outer
disks, in this case with the effect of overestimating the ratio of
$r_e^{blue}$ to $r_e^{disk}$). This value is quite consistent with
$r_e^{bulge}$ as estimated above.  However, the scatter in
$r_e^{blue}$ is large, and we have ignored the presence of underlying
disk color gradients, which are highly uncertain for blue-centered
galaxies.

Another uncertainty in assessing the starburst scale arises from
substructure within the starburst.  A few galaxies in Fig.~1 show
local reddening trends inside the shaded $r_e^{bulge}$ region,
superimposed on the general trend toward bluer colors with decreasing
radius.  Others show central blueing in excess of the general trend,
and some show more blueing in $U-B$ than in $B-R$ (Fig.~10), perhaps
indicating that dust has suppressed the blueing in $B-R$
\citep[although extinction is more severe in U and B than in R, it
affects $B-R$ colors more than $U-B$
colors;][]{gordon.calzetti.ea:dust}.  We speculate that this
substructure within the starburst color profiles may be related to the
differentiation of the bulge and the inner disk, via gradients in
starburst age, intensity, and/or dust formation.  Despite the
uncertainty in interpreting such features, we conclude that the
starbursts in blue-centered galaxies are plausibly related to bulge
formation.

\subsection{Externally Driven Bulge Growth without Disk Destruction}

All of the bulge growth we identify by the blue-center criterion
occurs within intact disk galaxies.  However, one may ask
whether these galaxies' disks will soon be destroyed by mergers.  Of the eleven
blue-centered galaxies in Fig.~1, only four have a larger neighbor
within 300 kpc and 300 \kms.  Only one of these, NGC~7752, appears to
be in direct contact with its larger companion.  Therefore major
mergers are unlikely for most of these galaxies for at least a few
Gyr.\footnote{In contrast, nearly all blue-centered galaxies in the
Close Pairs Survey have larger companions.  However, this fact appears
inseparable from the selection bias inherent in the survey: given that
blue-centered galaxies tend to have low luminosities, they will nearly
always be the junior companions in a magnitude-limited survey of
galaxy pairs.}  

We note that substantial gas inflow and bulge growth do not require large
companions.  \citet{hernquist.mihos:excitation} find that a 10:1 merger can
drive up to $\sim50\%$ of the {\em larger} galaxy's gas into its center, so
that the inflowing gas significantly outweighs the mass of later-accreted
material.  Minor mergers and interactions with small companions have been
observed to thicken but not destroy the disks of larger galaxies
\citep[e.g.,][]{schwarzkopf.dettmar:influence*1}.  Furthermore, subsequent
thin-disk regrowth is likely, occurring through both fresh gas accretion
and fall-back of tidal debris \citep[][]{barnes:formation}.

\subsection{Potential for Evolution in Bulge-to-Disk Ratio}
\label{sc:plausgrowth}

Although the blue colors and large H$\alpha$ equivalent widths of
blue-centered galaxies clearly reveal starburst activity, one may
nonetheless ask whether the starbursts are of sufficient intensity to
produce noticeable changes in bulge-to-disk ratio.  To answer this
question carefully would require H$\alpha$ imaging combined with
detailed modeling of the stellar populations and star formation
histories of blue-centered galaxies.  Such work is in progress and is
beyond the scope of this paper.  For the moment, we offer a simple
plausibility argument for the possibility of significant bulge growth
in blue-centered galaxies.

We consider an idealized three-component galaxy, with a preexisting
disk, a preexisting bulge, and a new bulge component that forms on top
of the old bulge during the blue-centered phase.  We construct a
population synthesis model for each component, treating early (S0--Sb)
and late ($>$Sb) Hubble types separately.  The early-type model has an
exponential disk, a coeval r$^{1/4}$-law bulge, and a new
r$^{1/4}$-law bulge.  The late-type model substitutes exponential
bulges.  The preexisting bulges follow $r_e^{old\,
bulge}=0.2r_e^{disk}$ and $r_e^{old\, bulge}=0.1r_e^{disk}$ for early
and late type galaxies, respectively \citep[][ and references
therein]{macarthur.courteau.ea:structure}, while the new bulges in
both models follow $r_e^{new\, bulge}=0.4r_e^{disk}$ (see discussion
in \S~\ref{sc:plausbulge}).

We use the spectral synthesis models of \citet[][; Salpeter IMF, solar
metallicity]{bruzual-a-.charlot:models} to generate the stellar
populations, adopting exponential star formation histories with a 1 Gyr
timescale for the preexisting bulge, a 4 Gyr timescale for the early-type
galaxy disk, and a 7 Gyr timescale for the late-type galaxy disk.  For the
new bulge, we consider two possible burst timescales, 100 Myr and 1 Gyr.
The shorter timescale represents a plausible lower limit based on detailed
modeling of externally triggered starbursts by
\citet{barton-gillespie.geller.ea:tidally}, while the longer timescale is
closer to the dynamical timescale and might encompass two or more
sub-bursts within a realistic encounter.  The new bulge starts to form 8
Gyr after the preexisting components.  Our models reproduce the average
central and outer disk colors of NFGS blue-centered galaxies, yielding
\dbr\ values similar to those observed (Fig.~11).  We match the effective
colors after 4 Gyr of fading to the colors of non-blue-centered galaxies in
the NFGS of early and late Hubble types.  Just before the new bulges form,
the preexisting bulges have bulge-to-total luminosity ratios $B/T$ $\sim$
0.2 and $\sim$ 0.04 for the early and late type models, respectively.
During the blue-centered phases, the models generally have appropriate
total $B/T$ for early and late type galaxies (combining the new+old
bulges), although the late-type model with the 100 Myr burst varies rapidly
in total $B/T$ from $\sim$0.6 down to $\sim$0.2 just within the brief
blue-centered phase.

Fig.~11 shows the time evolution of both \dbr\ and the new-bulge $B/T$.
The peak new-bulge $B/T$ reaches 0.5-0.7 for early types and 0.3--0.5 for
late types, representing substantial evolution in observed Hubble type
relative to the preexisting $B/T$ ratios of 0.2 and 0.04.  Even after 4 Gyr
of fading, the new bulge still contributes $B/T$ $\sim$ 0.1--0.2 for early
types and $\sim$0.04--0.07 for late types, implying fractional bulge growth
of $\sim$50--100\%.  We conclude that our simplified models support the
idea that blue-centered galaxies may experience substantial bulge growth,
and more detailed modeling would be valuable.

\subsection{Frequency of Blue-Centered Phases}
\label{sc:plausimp}

Blue-centered galaxies make up $\sim$10\% of disk galaxies in the NFGS
(Table~1).  Our population synthesis models suggest plausible durations for
the \dbr\ $>$ 0 phase of $\sim$0.5--2 Gyr for burst timescales of 100
Myr--1 Gyr (Fig.~11).  In reality the blue-centered phase will be shorter
to the extent that the starburst reddens as it develops dust.  From these
numbers alone, we estimate that in the last 10 Gyr, $\ga$50\% of disk
galaxies have experienced a blue-centered phase, and it is possible that
many disk galaxies have experienced two or more blue-centered phases.

However, we know that \dbr\ correlates with luminosity, and we have
suggested that large red preexisting bulges or bars in high-luminosity
galaxies may mask blue centers.  As expected, in the simplified models of
\S~\ref{sc:plausgrowth}, we find that blue-centered phases are suppressed
whenever the peak new-bulge $B/T$ ratio during the starburst is specified
to be smaller than the preexisting-bulge $B/T$ ratio.  Blue centers may
also be lost in the context of blue disks: our late-type blue-centered
galaxies have somewhat redder disks than the average late-type galaxy,
which must at least partly reflect selection bias.  We conclude that many
non-blue-centered galaxies may have experienced bulge growth by processes
similar to those acting in blue-centered galaxies, and the data are {\em
consistent} with a hierarchical scenario in which most disk galaxies
experience repeated {\em in situ} bulge growth phases driven by
interactions and minor mergers.

\section{Open Questions}
\label{sc:openq}

The foregoing analysis leaves two tantalizing questions unanswered.  First,
are the young bulges forming in blue-centered galaxies likely to be disky?
Second, how do blue-centered galaxies fit into hierarchical galaxy
formation scenarios?

\subsection{Do blue-centered galaxies form disky bulges?}

Sersic $n$ analysis is infeasible with the present data and sample, so we
cannot determine whether our blue-centered galaxies are forming exponential
bulges.  However, we can use the Faber-Jackson relation to search for
evidence of disky $v/\sigma$ ratios, in the spirit of
\citet{kormendy:kinematics}.  Fig.~12a shows the Faber-Jackson relation for
elliptical galaxies and bulges in the NFGS, measuring bulge luminosities
from ``any light in excess of an inward extrapolation of a constant
scale-length exponential disk'' \citep{wyse.gilmore.ea:galactic}, where the
exponential disk profiles come from the outer-disk fits in
\S~\ref{sc:plausbulge}.  Only four blue-centered galaxy bulges are shown,
as the others either lack velocity dispersion data or have M$_{\rm B}$ or
$\sigma$ measurement errors comparable to the plot scale \citep[note that
the instrumental resolution in most cases is $\sigma\sim60$
\kms,][]{kannappan.fabricant:broad}. These four blue-centered galaxy bulges
are all offset toward low $\sigma$ in the Faber-Jackson relation,
consistent with disky kinematics as discussed by
\citet{kormendy:kinematics}.  However, Fig.~12b shows that their offsets
fall within a broad correlation between color and Faber-Jackson residuals.
The slope of this correlation is consistent with stellar population effects
\citep[by analogy with the arguments in][ regarding the correlation between
color and residuals from the Tully-Fisher
relation]{kannappan.fabricant.ea:physical}.  Any kinematic diskiness, if
present, is hidden in the noise of the color trend.

Nonetheless, the possibility that externally triggered {\em in situ} bulge
growth may produce disky bulges merits further scrutiny, for two reasons.
First, the three prototype kinematically disky bulges discussed by
\citet{kormendy:kinematics} all show possible evidence of interactions and
mergers as well as secular evolution. NGC~4736 has a double nucleus in HST
UV imaging and may be a merger remnant \citep{maoz.filippenko.ea:detection,
shioya.tosaki.ea:molecular}; NGC~4826 shows counterrotation consistent with
a 10:1 merger \citep[stars vs.\ gas and gas vs.\
gas,][]{rix.kennicutt.ea:placid}; and NGC~7457 has a companion in NED with
$\Delta X \sim 30$ kpc and $\Delta V \sim 100$ \kms.  The role of galaxy
encounters in shaping these galaxies is an open question.

Second, unlike the collisionless simulations of
\citet{aguerri.balcells.ea:growth}, the hydrodynamic simulations of
\citet{scannapieco.tissera:effects} show that interactions and even mergers
can decrease Sersic $n$.  Gas inflow and star formation during the
pre-merger orbital decay phase tends to decrease Sersic $n$, while the
merger that follows rarely completely reverses the
change.\footnote{\label{fn:secword} Note that Scannapieco \& Tissera use
the term ``secular'' to refer to interaction-driven gas inflow, differing
from usage in this paper.}  Within a hierarchical simulation volume 5$h^{-1}$
Mpc on a side, Scannapieco \& Tissera find that most bulges show Sersic
$n\la1$ by $z=0$.  Further simulations would be valuable to verify these
results for larger samples of simulated galaxies.

\subsection{How do blue-centered galaxies fit into hierarchical galaxy formation scenarios?}

\citet{scannapieco.tissera:effects} present a picture in which interactions
and minor mergers drive an evolutionary loop between early and late type
disk galaxies, somewhat analogous to the loop between elliptical galaxies
and disk galaxies seen in lower resolution simulations.  As interactions
and mergers reshape galaxy bulges, and new thin disks accrete over the
thick disks created by previous encounters, at any epoch we observe a
snapshot Hubble sequence \citep[see
also][]{steinmetz.navarro:hierarchical}.

This picture offers an alternative to the bulge rejuvenation scenario
proposed by \citet{ellis.abraham.ea:relative}.  Using the Hubble Deep
Field, Ellis et al.\ show that bulges are bluer than elliptical galaxies at
every epoch to $z\sim1$, contradicting simplified hierarchical models in
which bulges represent old elliptical galaxies around which new disks have
grown.  The authors find evidence that bulges have been rejuvenated in
short bursts of star formation, and they suggest that this rejuvenation may
reflect internal secular processes.  We suggest instead that both the
bursty rejuvenation Ellis et al.\ observe at high $z$ and the blue-centered
phases we observe at low $z$ reflect ``high-resolution'' hierarchical
galaxy formation, involving minor companions and small-scale, local gas
physics.  In fact, the tabulated data of Ellis et al.\ indicate a high rate
of galaxies with bluer central colors than their overall colors ($\sim$30\%
of the spiral sample described in their Figure~4), with the caveat that
most of the blue excesses do not exceed the errors (R. Abraham, private
communication).

The luminosity distribution of blue-centered galaxies at both low and high
$z$ may reflect hierarchical processes.  At the current epoch, most
blue-centered galaxies are fainter than M$_{\rm B}\sim-20$.  This trend may
be related to the hierarchical tendency for the faint end of the luminosity
function to show continued galaxy formation activity at later epochs than
does the bright end \citep{cowie.songaila.ea:new}, and luminosity-dependent
color masking (\S~\ref{sc:driversa}) may be a corollary effect.  In this case,
bright galaxies would have experienced their blue-centered phases at higher
redshift, at a time when they enjoyed both a higher merger rate and a
greater starburst efficiency, because of larger gas reservoirs. Luminous
blue compact galaxies \citep[LBCGs, e.g.,][]{guzman.koo.ea:on} seen at high
$z$ may represent exactly this phenomenon.  LBCGs have been interpreted as
bulges in formation by \citet{kobulnicky.zaritsky:chemical} and
\citet{barton.:possible}, though their exact nature is still an open
question.

At low $z$, analogues of high-$z$ LBCGs often take the form of
low-luminosity emission-line S0 galaxies (see the discussion of Barton \&
van Zee 2001 in Kannappan \& Barton 2003). Four of the eleven blue-centered
galaxies in the NFGS are emission-line S0 galaxies (including type S0/a,
Fig.~2), in keeping with the general abundance of S0 galaxies at low
luminosities.  Many non-blue-centered low-luminosity S0 galaxies also show
central starburst activity and evidence of mergers or interactions
\citep[e.g., counterrotating gas and stars,][ and references
therein]{kannappan.fabricant:broad}.  We speculate that unlike bright S0
galaxies in high-density environments, low-luminosity S0 galaxies may often
represent transient objects, caught at one extreme of the morphological
evolutionary loop described by Scannapieco \& Tissera.  If so, then these
recently formed bulge+thick disk systems may be destined to fade, regrow
thin disks, and finally fill in the low-luminosity spiral sequence.

\section{Conclusion}
\label{sc:concl}

Recent research on young and/or ``disky'' bulges
\citep[e.g.,][]{kormendy:kinematics,andredakis.peletier.ea:shape,peletier.balcells:ages,courteau..ea:evidence,carollo.stiavelli.ea:hubble}
has pointed to the importance of late-epoch bulge formation within
preexisting disks ({\em ``in situ''} bulge formation).  However, this
body of work has not resolved an interpretive degeneracy between {\em
in situ} formation scenarios involving spontaneous disk instabilities
and those involving externally triggered instabilities.  As a first
step toward breaking this degeneracy, we have identified a class of
galaxies likely to be experiencing active {\em in situ} bulge growth,
based on the observation that their centers are bluer than their outer
disks \citep[as
in][]{schweizer:interactions,barton-gillespie.geller.ea:tidally}.  To
determine whether the primary drivers of the blue-centered phenomenon
are external or internal, we have examined the properties and
environments of blue-centered galaxies in the Nearby Field Galaxy
Survey
\citep[NFGS,][]{jansen.fabricant.ea:spectrophotometry,jansen.franx.ea:surface,kannappan:kinematic},
a statistically representative sample of the low-redshift galaxy
population.  The NFGS contains eleven blue-centered galaxies
(excluding dwarf galaxies and AGN), all of which are star-forming disk
galaxies, with no preferred Hubble type or global environment.

We find that blue-centered galaxies correlate at 95--99\% confidence
with two types of evidence of external disturbance: (1) morphological
peculiarities, and (2) nearby companions in the Updated Zwicky Catalog
\citep[UZC,][]{falco.kurtz.ea:updated}.  The peculiarity correlation
is independent of the separate tendency for blue-centered galaxies to
have low luminosities
\citep[][]{tully.verheijen.ea:ursa,jansen.franx.ea:surface}, and in
fact the peculiarity correlation strengthens to 99.5\% confidence if
we correct the color difference parameter \dbr\ (outer-disk color
minus color within the half-light radius) for the luminosity--\dbr\
trend.  However, the UZC companion--\dbr\ correlation may not be
independent of the luminosity--\dbr\ trend, because of the fact that
lower luminosity galaxies are more likely to have companions in the
UZC.  On the other hand, many of the apparent companions that seem
most likely to be interacting with the blue-centered galaxies in our
sample are not listed in the UZC, because these companions are too
faint, too close, or already merging.  Including this type of
evidence, we find signs of possible interactions or mergers for all
eleven of the blue-centered galaxies in our study.

We have presented four quantitative plausibility arguments to link
these results more directly to the phenomenon of {\em in situ} bulge
growth.  First, the spatial scales of the starbursts in blue-centered
galaxies are consistent with the spatial scales expected for bulge
growth, especially taking into account the likelihood that bulges and
inner disks evolve together \citep{peletier.balcells:ages}.  Second,
most blue-centered galaxies in the NFGS appear to be merging or
interacting with smaller galaxies, consistent with bulge growth
mechanisms that heat but do not destroy disks.  Third, schematic
population synthesis models tuned to the observed properties of early-
and late-type blue-centered galaxies in the NFGS show significant
bulge growth, with fractional growth $\sim$50--100\%. And
fourth, a duty cycle argument based on the frequency of blue-centered
galaxies ($\sim$10\% of disk galaxies in the NFGS) and a plausible
burst timescale (100 Myr--1 Gyr) implies that the majority of disk
galaxies have experienced blue-centered phases, possibly more than
once during their lifetimes.  Although we observe blue-centered phases
primarily in galaxies fainter than M$_{\rm B}=-20$, bright galaxies
may well be experiencing almost-blue-centered phases, particularly if
the tendency for blue-centered galaxies to be faint simply reflects
greater color masking in bright galaxies (i.e., greater dilution of
the central blue light by preexisting red bulge populations).

These arguments support the idea that blue-centered phases represent
an important mode of {\em in situ} bulge growth in the life cycles of
most disk galaxies, driven primarily by interactions and mergers with
smaller companions.  Qualitatively, this scenario fits naturally
within the framework of hierarchical galaxy formation, and in fact
{\em in situ} bulge growth has been observed in high-resolution
hierarchical simulations that include small companions and local gas
dynamics
\citep[][]{tissera.domnguez-tenreiro.ea:double,steinmetz.navarro:hierarchical}.
The hierarchical framework may offer a deeper explanation for the
tendency of blue-centered galaxies to be faint: the luminosity trend
may reflect the mass dependence of galaxy formation timescales
\citep{cowie.songaila.ea:new}.  If so, then more massive galaxies
should have experienced blue-centered phases at earlier epochs.  This
prediction agrees qualitatively with evidence for ``bulge
rejuvenation'' in bright galaxies at high $z$
\citep{ellis.abraham.ea:relative}, and may also be related to the
phenomenon of luminous blue compact galaxies
\citep[e.g.,][]{guzman.koo.ea:on,kobulnicky.zaritsky:chemical,barton.:possible}.
The hierarchical paradigm may even offer a way to make disky bulges:
\citet{scannapieco.tissera:effects} find that pre-merger gas inflow
processes within high-resolution hierarchical simulations help to
reshape galaxies toward exponential bulge profiles.  Observational
studies to date cannot confirm or reject a connection between
interactions and disky bulges.  In the NFGS, blue-centered galaxies
show offsets from the Faber-Jackson relation that are consistent with
either disky kinematics or young stellar populations.  Evaluating the
respective roles of hierarchical and internal secular processes in
forming the full variety of bulges in the general galaxy population
remains a challenge for future research.

\acknowledgments

We thank John Kormendy for valuable discussions on secular evolution,
and Bob Abraham for useful feedback on our population synthesis
models.  Margaret Geller, Rob Kennicutt, and Karl Gebhardt provided
helpful comments on the manuscript.  EJB acknowledges support from
NASA through Hubble Fellowship grant HST-HF-01135.01, awarded by
STScI, which is operated by AURA for NASA under contract NAS 5-26555.
This research has used the NASA/IPAC Extragalactic Database (NED),
operated by the Jet Propulsion Laboratory, California Institute of
Technology (Caltech) under contract with NASA.  This research has also
used NASA's Astrophysics Data System Bibliographic Services, as well
as data products from the Two Micron All Sky Survey, which is a joint
project of the University of Massachusetts and the Infrared Processing
and Analysis Center/California Institute of Technology, funded by the
National Aeronautics and Space Administration and the National Science
Foundation.\\

\appendix

\section{Notes on Individual Objects}

A10504+0454 has no UZC companion, but deep high-resolution imaging
(Kannappan, Impey, \& Mathis, in prep.) reveals a possible interacting
satellite, undetected in the image shown in Fig.~1.

A11332+3536 has non-coplanar counterrotating gas and stars
\citep{kannappan.fabricant:broad}.

A12001+6439 appears smooth at the resolution of Fig.~1, but deep
high-resolution imaging (Kannappan, Impey, \& Mathis, in prep.) reveals
loops and filaments suggestive of a late-stage merger.

NGC~3846 has no UZC companion, but its image in Fig.~1 shows a trail of
debris and two bright knots on the south side suggestive of a merging
companion.

NGC~5541 has no UZC companion, but its K-band image from the 2MASS All-Sky
Data Release reveals that the peculiarity seen at the top of the B-band
image (Fig.~1) is actually another galaxy, apparently interacting with
NGC~5541.  High-resolution radio imaging from the VLA FIRST survey
\citep{becker.helfand.ea:first} shows strong emission both in the companion
bulge and outside the primary bulge, but not within the primary bulge.  The
primary bulge is much redder than the bulge of any other blue-centered
galaxy and has an H$\alpha$ equivalent width of only 6.5\AA\ in emission,
despite a very blue region immediately surrounding the red core (Fig.~1)
and integrated EW(H$\alpha$) $\sim$ 6.5 \AA\ in emission.  Thus although
its \dbr\ identifies NGC~5541 as a rare high-luminosity blue-centered
galaxy, its low-luminosity companion probably has more similarities to the
rest of our blue-centered galaxy sample than NGC~5541 does.

NGC~5875A has no UZC companion, but its 2MASS color-composite image reveals
a large region with distinct JHK-color on the northeast side of the galaxy,
resembling a merging companion.

NGC~7752 appears to be in direct contact with the tidally distorted spiral
arm of its larger companion, NGC~7753.  The arm is not easily visible at
the scale and contrast of Fig.~1.

\newpage


\vspace{0.6in}

\noindent\leavevmode
\noindent\makebox[\textwidth]{
   \centerline{
      \psfig{file=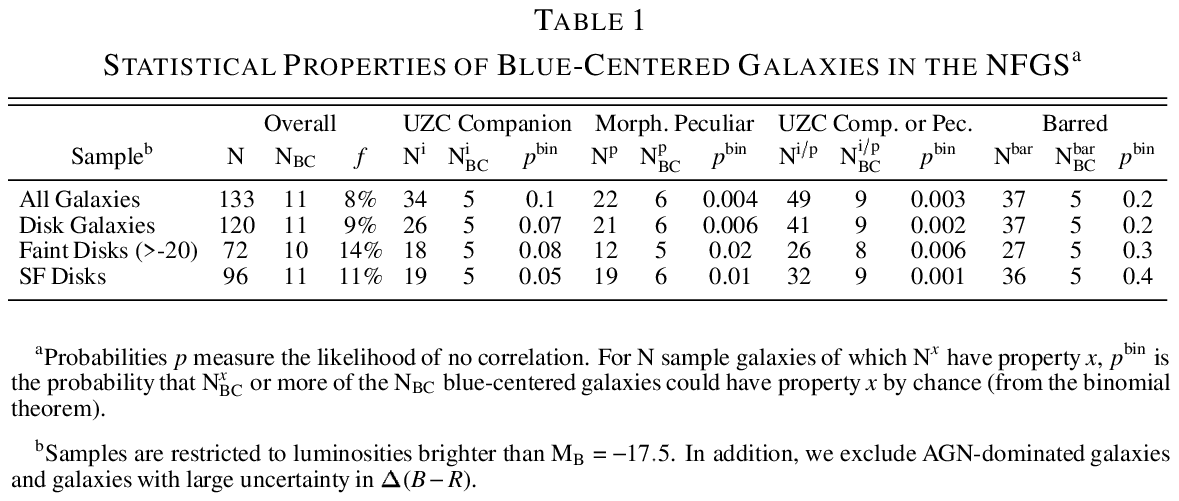,width=0.8\textwidth,clip=}
   }
}\par
\noindent\makebox[\textwidth]{
   \centerline{
\parbox[t]{\textwidth}{} }
}


\newpage
\noindent\leavevmode
\noindent\makebox[\textwidth]{
   \centerline{
      \psfig{file=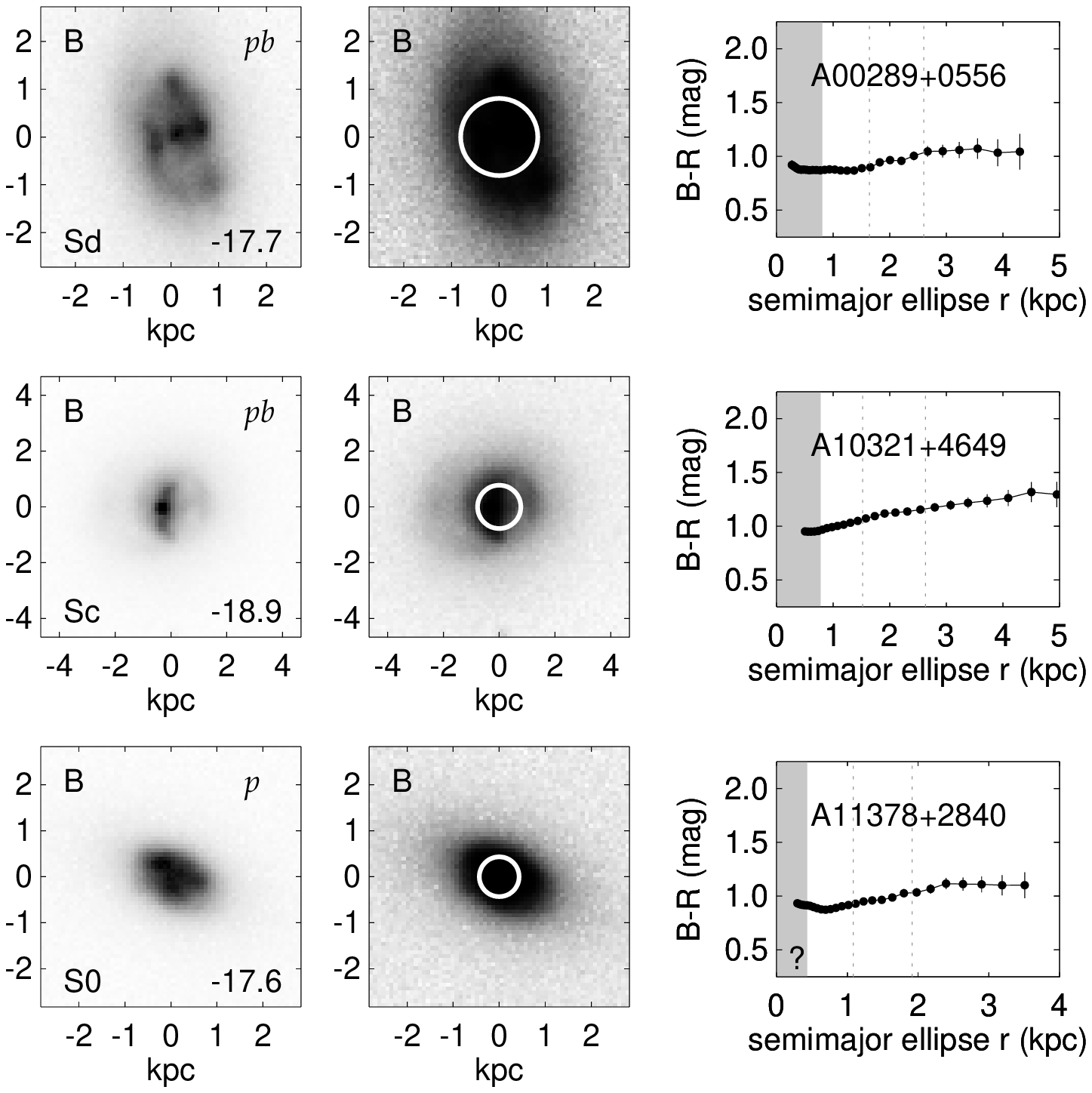,width=0.85\textwidth,clip=}
   }
}\par
\noindent\makebox[\textwidth]{
   \centerline{
\parbox[t]{\textwidth}{\footnotesize {\sc Fig.~1 ---} Images and radial
color profiles for the 11 blue-centered galaxies brighter than $M_{\rm
B}=-17.5$ in the NFGS.  Images are shown at two different contrast settings
except for NGC~5541, for which we show similar contrast settings in B and
K.  Optical data come from J00, and K-band data come from the 2MASS All-Sky
Data Release postage stamp server.  Italic labels $u$, $p$, and $b$
indicate close companions in the UZC, morphological peculiarity flags, and
bars, respectively.  Some galaxies show non-UZC companions, but these do
not receive a ``u'' label.  All distances are in kpc, with radii in the
color profiles expressed as major axis distances.  Vertical dashed lines
mark the half-light and 75\%-light radii in each color profile.  Central
colors influenced by seeing effects are not plotted.  The shaded regions of
the color profiles and the white circles on the images show the bulge
half-light regions one might plausibly expect for the eventual relaxed
bulge+disk systems, assuming minimal change in the outer-galaxy light
profiles (see discussion in \S~\ref{sc:plausbulge}).  Question marks in
some of the shaded regions indicate uncertain estimates.} } }\\

\newpage

\noindent\leavevmode
\noindent\makebox[\textwidth]{
   \centerline{
      \psfig{file=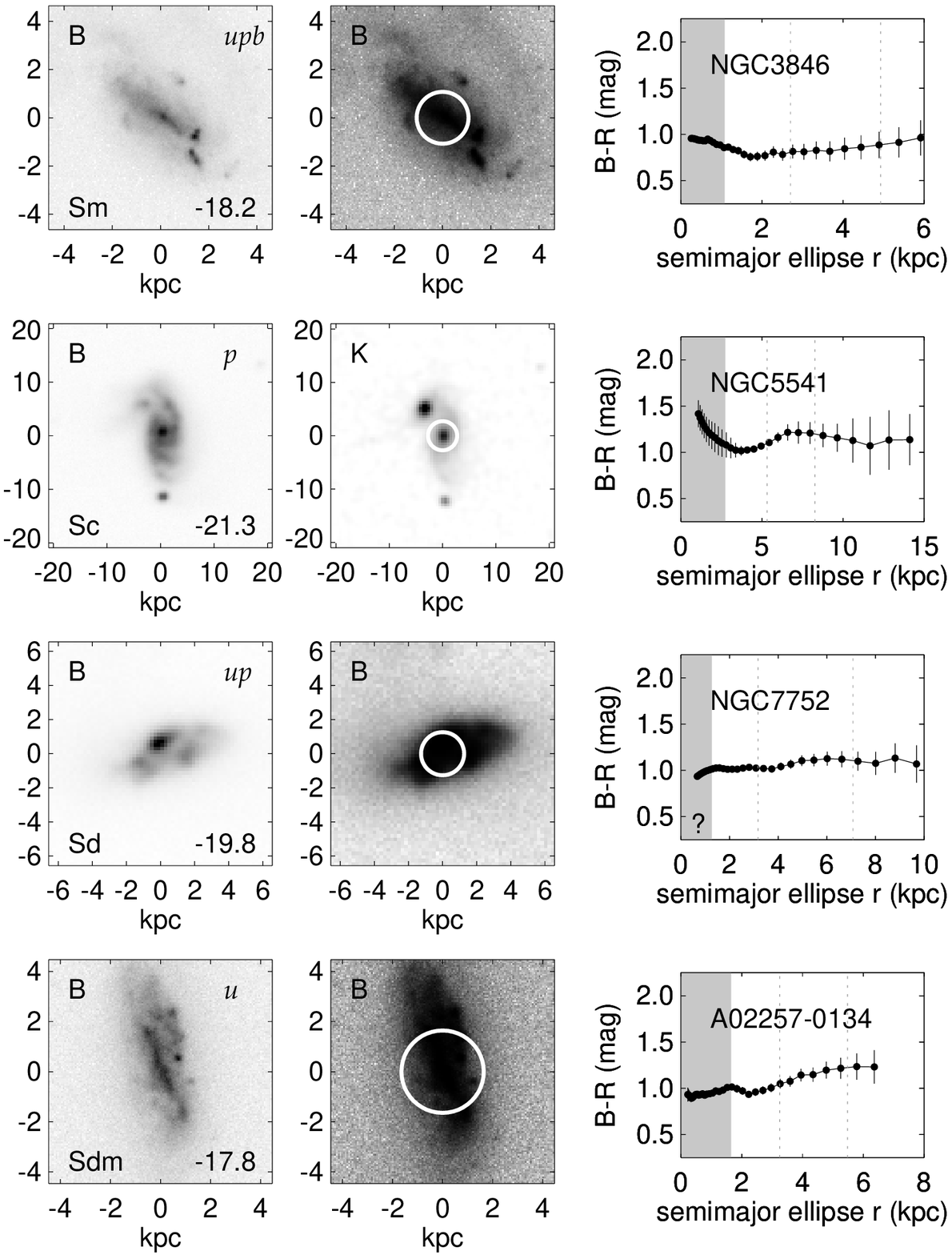,width=0.9\textwidth,clip=}
   }
}\par
\noindent\makebox[\textwidth]{
   \centerline{
\parbox[t]{\textwidth}{\footnotesize {\sc Fig.~1 ---} {\em Continued}
} } }\\

\newpage

\noindent\leavevmode
\noindent\makebox[\textwidth]{
   \centerline{
      \psfig{file=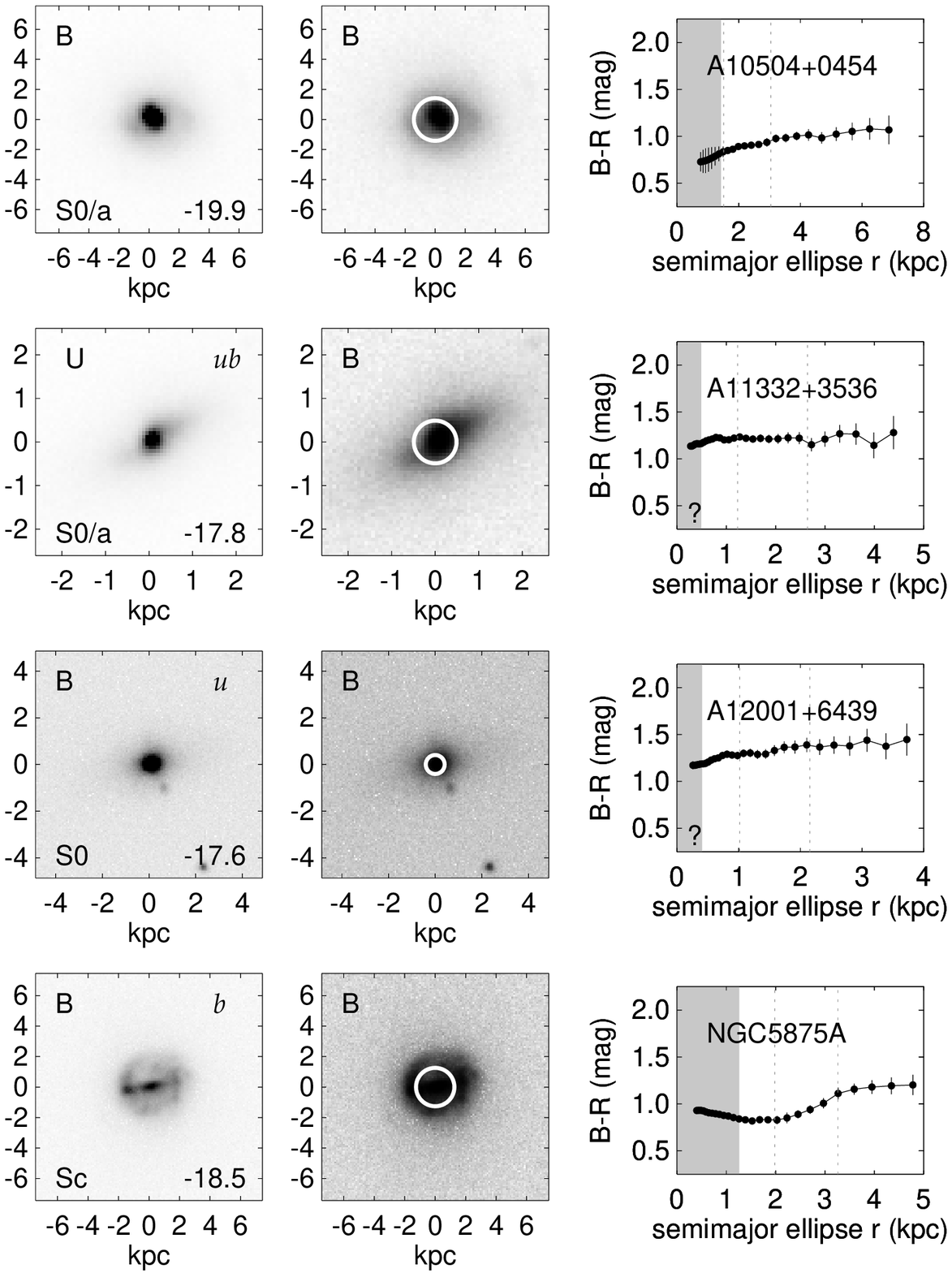,width=0.9\textwidth,clip=}
   }
}\par
\noindent\makebox[\textwidth]{
   \centerline{
\parbox[t]{\textwidth}{\footnotesize {\sc Fig.~1 ---} {\em Continued}
} } }\\


\noindent\leavevmode
\noindent\makebox[\textwidth]{
   \centerline{
      \psfig{file=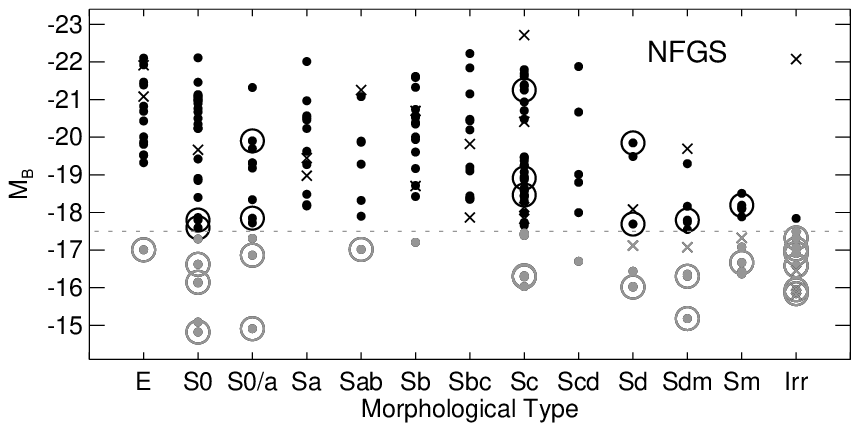,width=0.8\textwidth,clip=}
   }
}\par
\noindent\makebox[\textwidth]{
   \centerline{
\parbox[t]{\textwidth}{\footnotesize {\sc Fig.~2 ---} Distribution of
absolute magnitude and morphological type for the full NFGS sample.
Circles identify blue-centered galaxies.  The dashed line shows the lower
luminosity cutoff for the sample of 133 galaxies analyzed in this paper
(\S~\ref{sc:bcid}).  Gray symbols mark galaxies excluded by the luminosity
cut, while crosses mark galaxies excluded from analysis because of large
errors in \dbr\ or dominant AGN (\S~\ref{sc:bcid}).} } }\\



\noindent\leavevmode
\noindent\makebox[\textwidth]{
   \centerline{
      \psfig{file=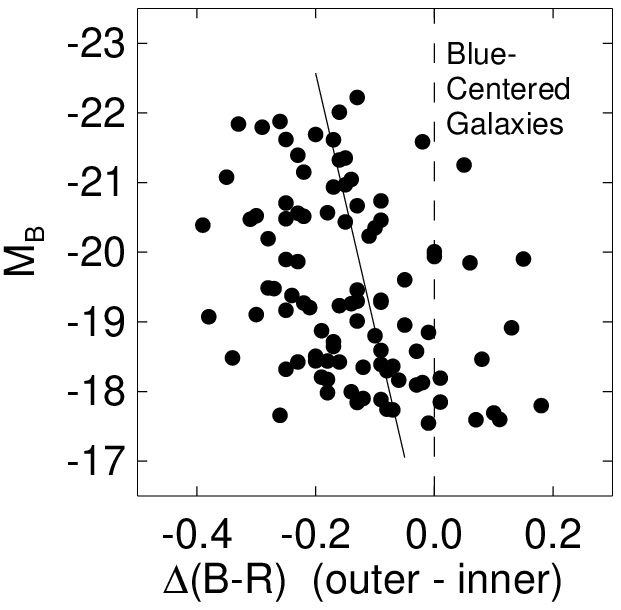,width=0.4\textwidth,clip=}
      \psfig{file=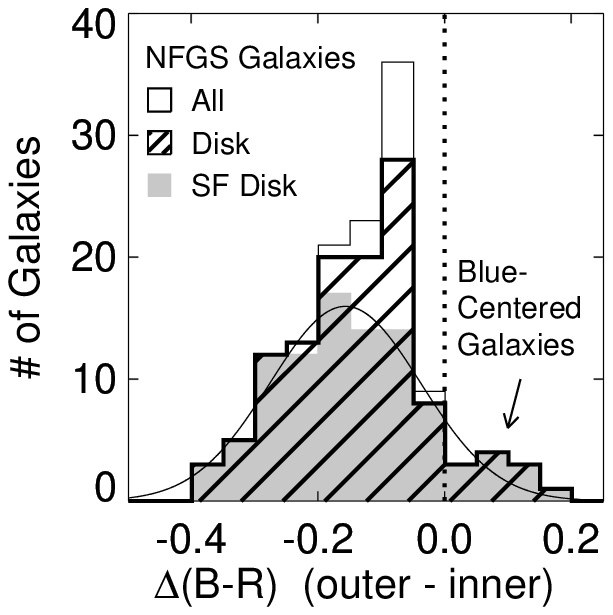,width=0.4\textwidth,clip=}
   }
}\par
\noindent\makebox[\textwidth]{
   \centerline{
\parbox[t]{\textwidth}{\footnotesize {\sc Fig.~3 ---} (\emph{a}) Luminosity
dependence of \dbr\ for star-forming disk galaxies in the NFGS. The line is
a least-squares inverse fit (\dbr\ as a function of luminosity) with
equation M$_{\rm B}=-15.2+30.7\times\dbr$.  (\emph{b}) Distribution of
\dbr\ for NFGS galaxies of all morphological types (open histogram), with
sub-distributions for disk galaxies (hashed) and star-forming disk galaxies
(shaded).  The Gaussian fit to the star-forming disk galaxy sample has mean
$-0.16$ and $\sigma=0.12$~mag.  Both panels and all remaining figures
exclude galaxies fainter than M$_{\rm B}=-17.5$ and galaxies marked with a
cross in Fig.~2.}  } }\\


\noindent\leavevmode
\noindent\makebox[\textwidth]{
   \centerline{
      \psfig{file=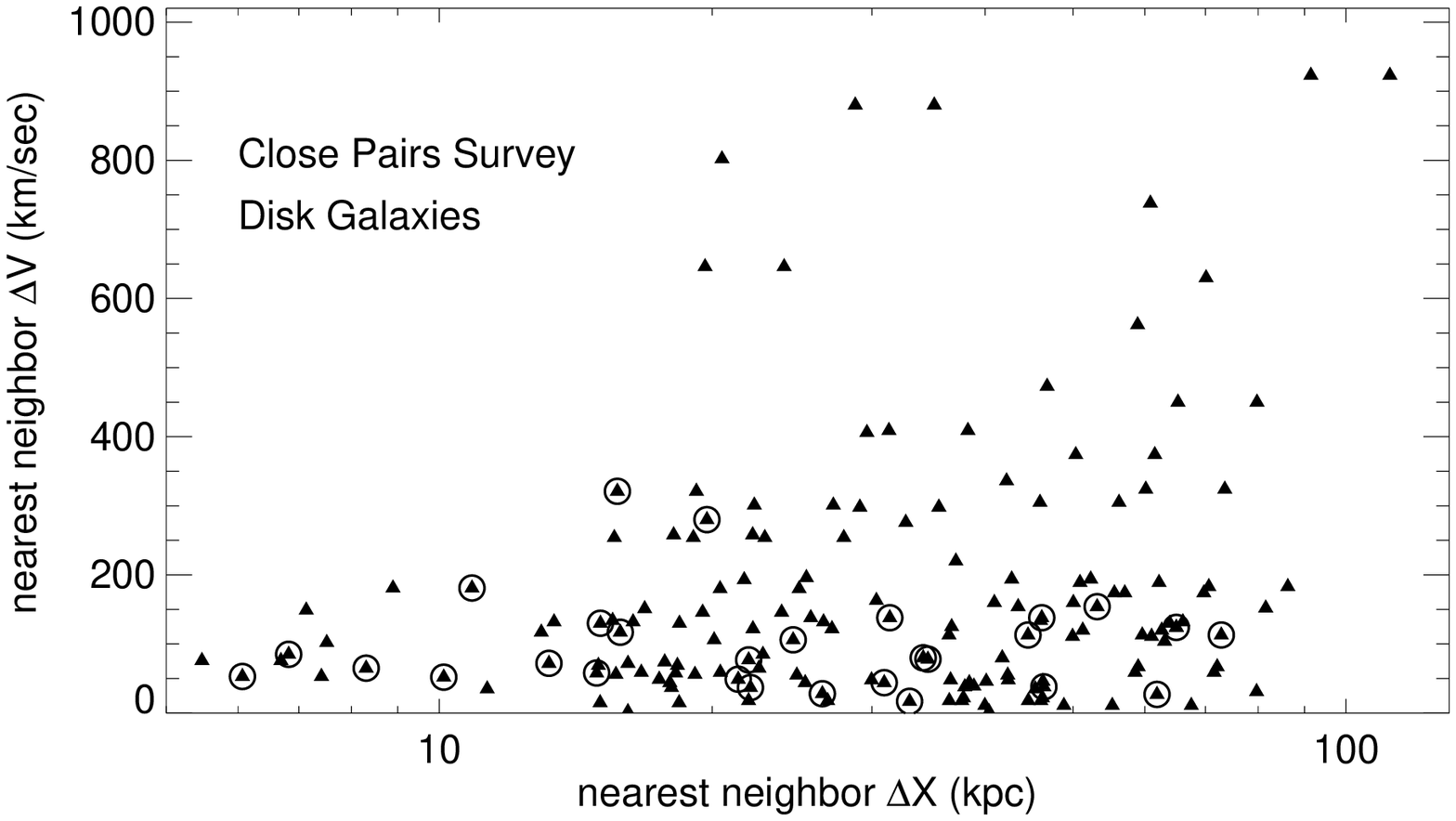,width=0.8\textwidth,clip=}
   }
}\par
\noindent\makebox[\textwidth]{
   \centerline{
\parbox[t]{\textwidth}{\footnotesize {\sc Fig.~4 ---} Projected distance
and line-of-sight velocity separation to the nearest neighbor for galaxies
in the Close Pairs Survey.  Blue-centered galaxies are circled. Separate
K-S tests on $\Delta V$ and $\Delta X$ indicate that blue-centered galaxies
tend to have smaller neighbor separations than non-blue-centered galaxies
at 97\% and 95\% confidence, respectively. } } } \\


\noindent\leavevmode
\noindent\makebox[\textwidth]{
   \centerline{
      \psfig{file=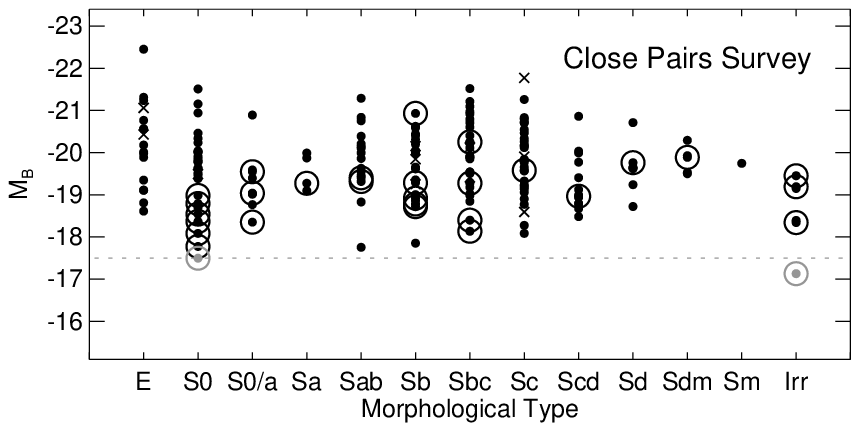,width=0.8\textwidth,clip=}
   }
}\par
\noindent\makebox[\textwidth]{
   \centerline{
\parbox[t]{\textwidth}{\footnotesize {\sc Fig.~5 ---} Luminosity
and morphology distribution for the full Close Pairs Survey.  Circles
identify blue-centered galaxies.  Crosses mark galaxies excluded from
analysis because of large photometric errors or dominant AGN.  The
dashed line shows the lower luminosity cutoff for the sample analyzed
in this paper (\S~\ref{sc:bcid}). Gray symbols mark galaxies excluded
by the luminosity cut.} } }\\



\noindent\leavevmode
\noindent\makebox[\textwidth]{
   \centerline{
      \psfig{file=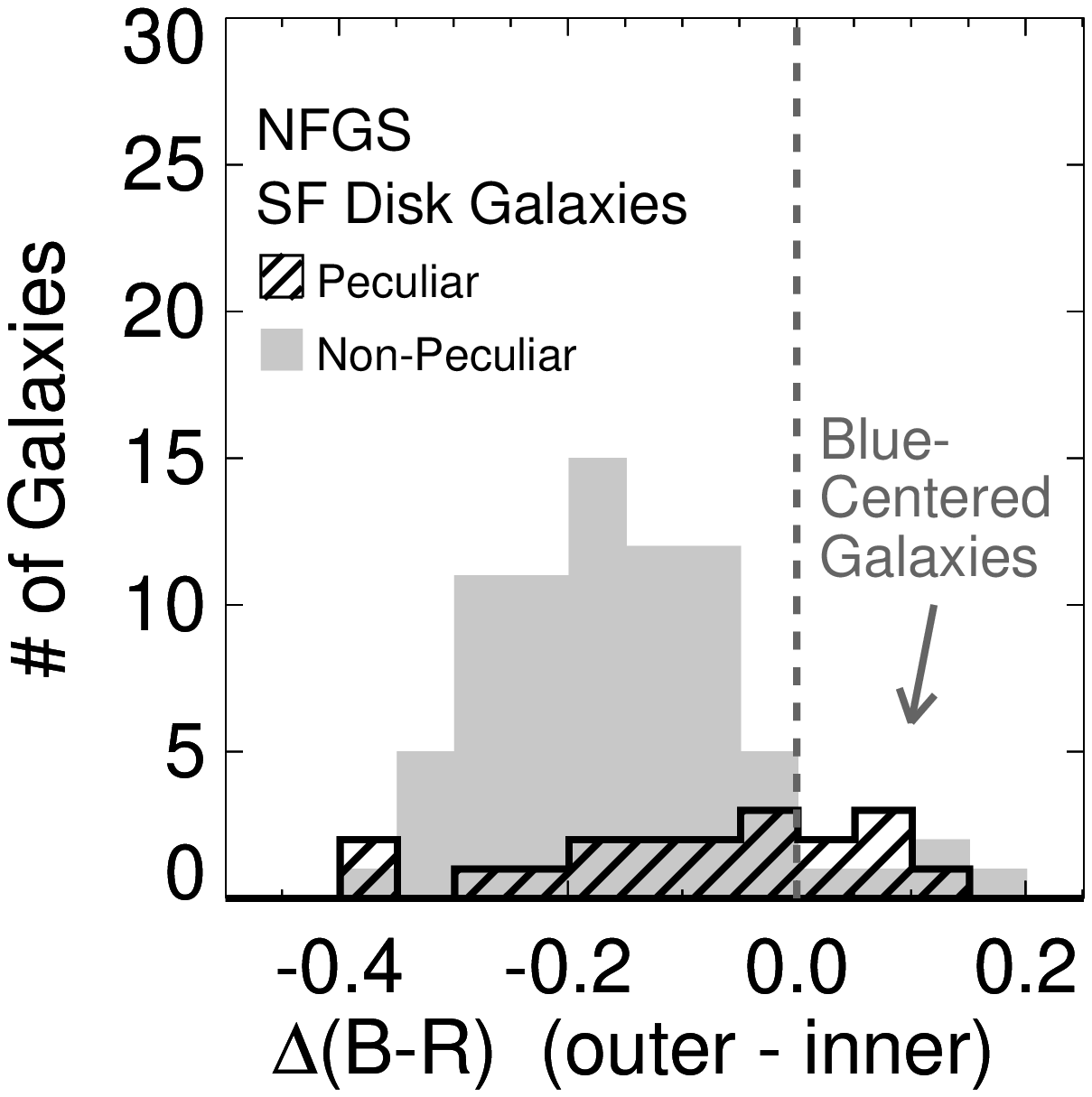,width=0.4\textwidth,clip=}
      \psfig{file=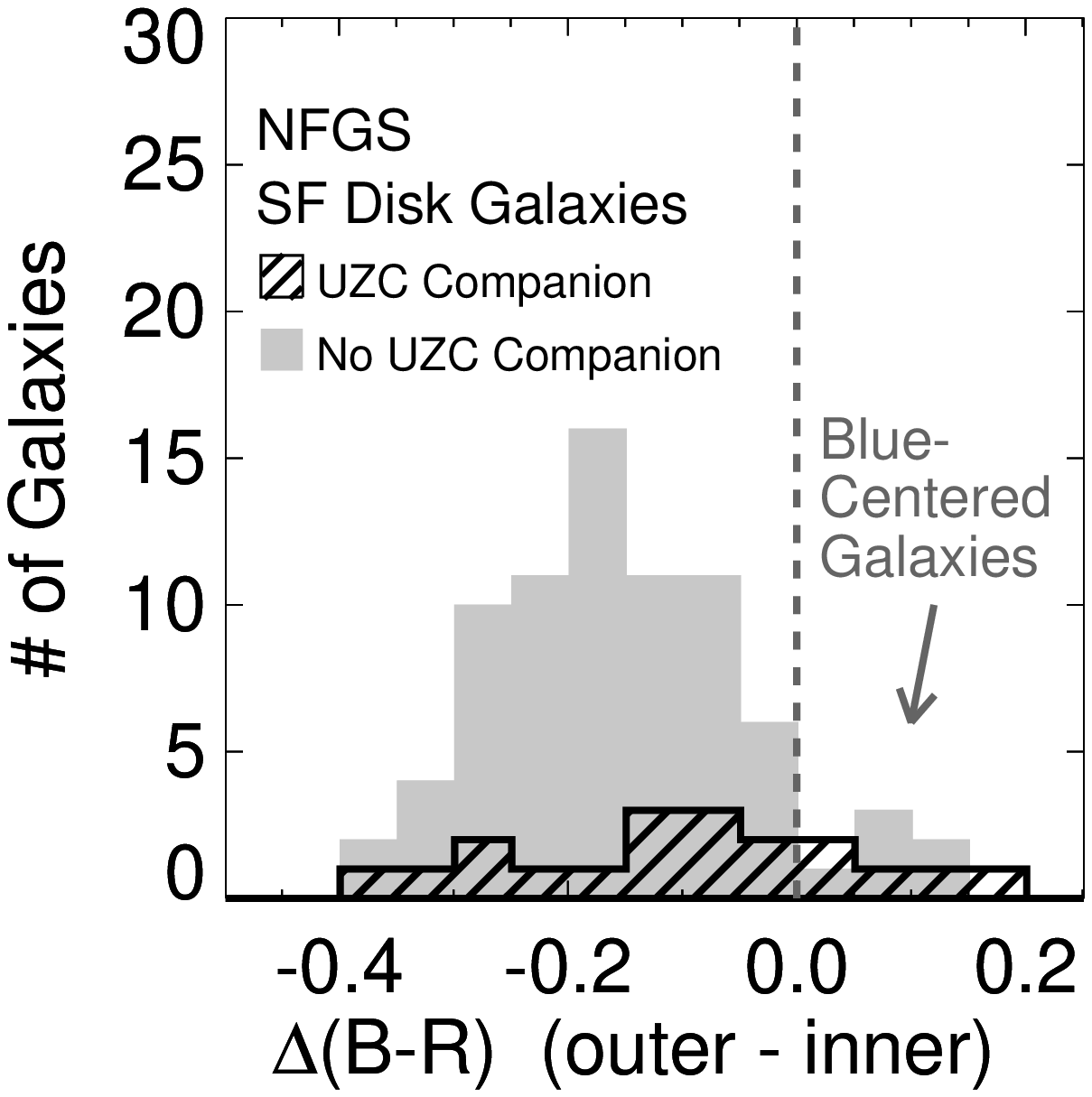,width=0.4\textwidth,clip=}
   }
}\par
\noindent\makebox[\textwidth]{
   \centerline{
\parbox[t]{\textwidth}{\footnotesize {\sc Fig.~6 ---} (Distribution of the
color difference \dbr\ for star-forming disk galaxies \emph{a}) with and
without morphological peculiarities, and (\emph{b}) with and without close
companions in the Updated Zwicky Catalog
\citep[UZC,][]{falco.kurtz.ea:updated}.  } } }\\



\noindent\leavevmode
\noindent\makebox[\textwidth]{
   \centerline{
      \psfig{file=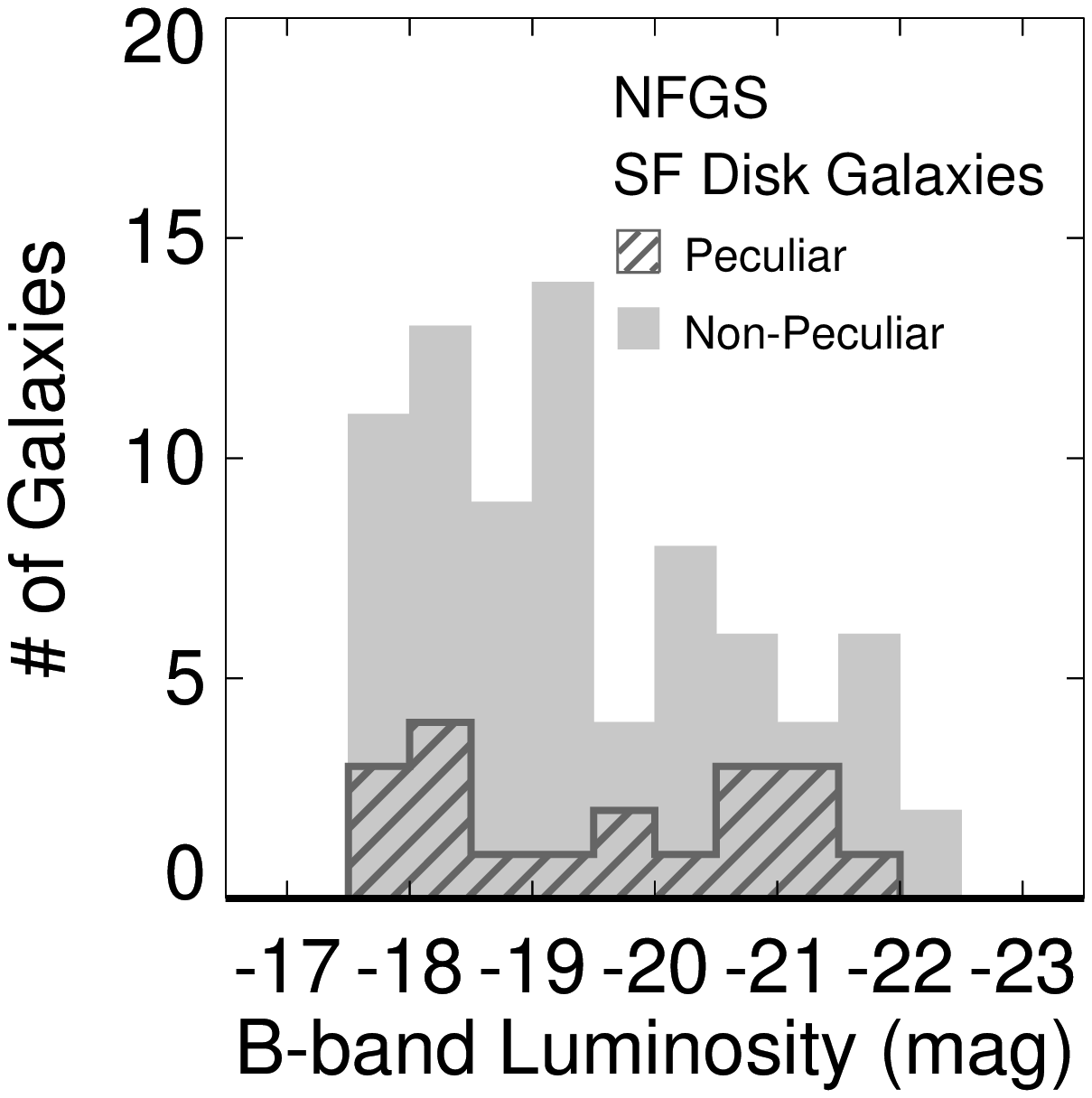,width=0.4\textwidth,clip=}
      \psfig{file=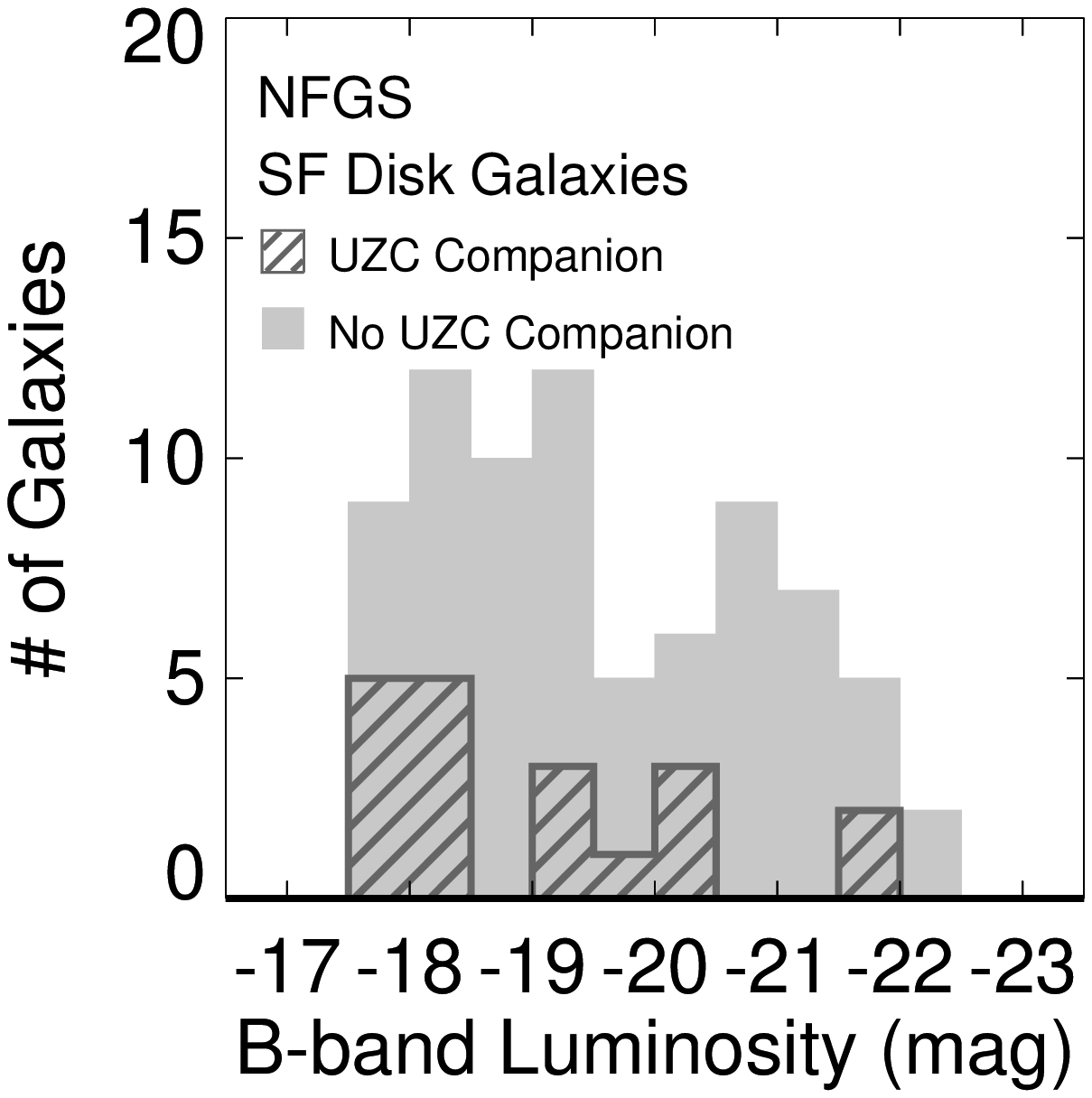,width=0.4\textwidth,clip=}
   }
}\par
\noindent\makebox[\textwidth]{
   \centerline{
\parbox[t]{\textwidth}{\footnotesize {\sc Fig.~7 ---} Luminosity distributions for star-forming disk galaxies (\emph{a}) with and
without strong morphological peculiarities, and (\emph{b}) with and without
close companions in the UZC.  Note that the tendency for fainter galaxies
to have more companions does not reflect distance-dependent sampling bias,
but rather the fact that the luminosity function cuts off at high
luminosities (\S~\ref{sc:samples} \& \ref{sc:driversa}).  } } }\\


\newpage
\noindent\leavevmode
\noindent\makebox[0.485\textwidth]{
   \centerline{
      \psfig{file=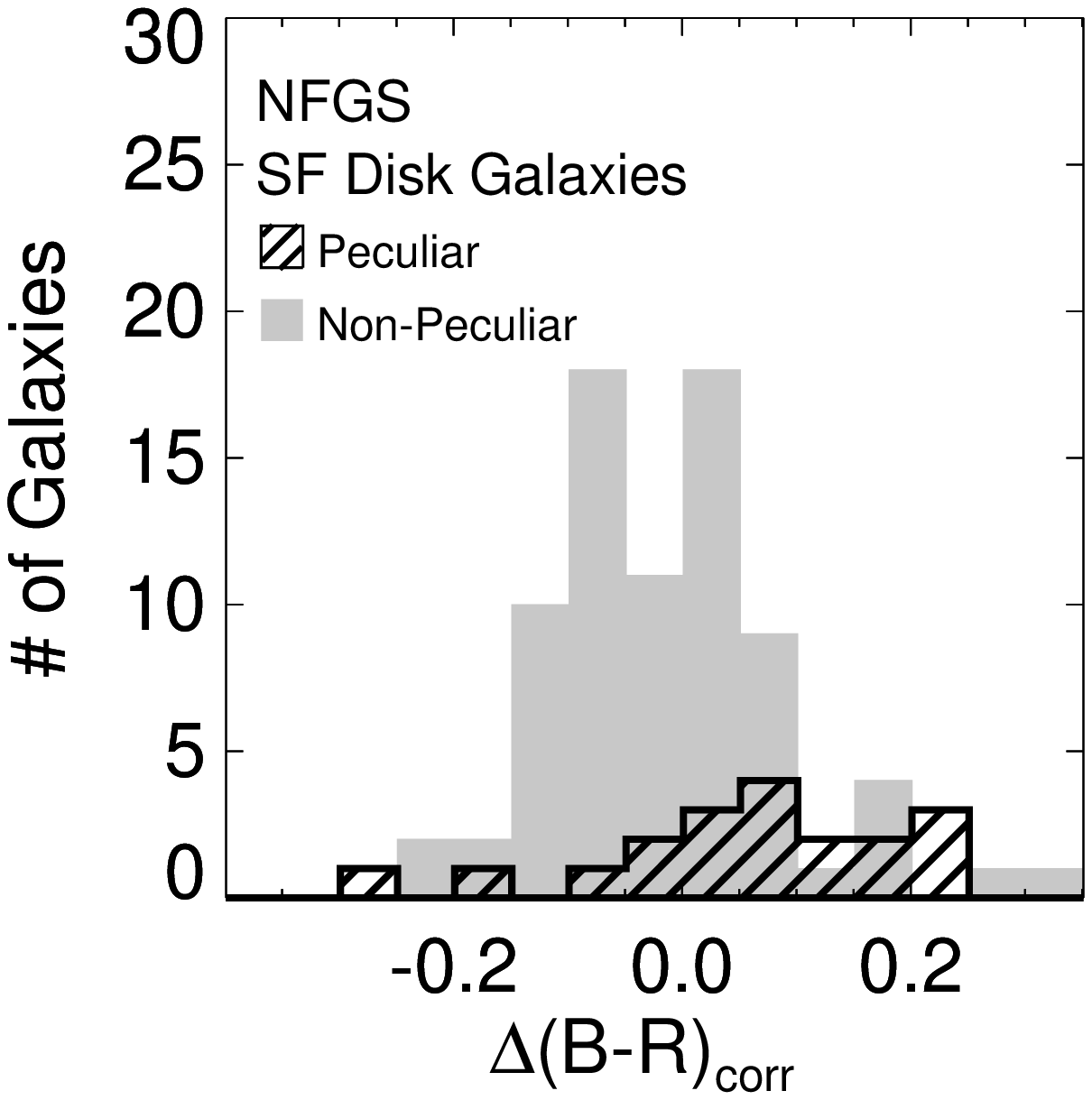,width=0.42\textwidth,clip=}
   }
}\par
\noindent\makebox[0.485\textwidth]{
   \centerline{
\parbox[t]{0.485\textwidth}{\footnotesize {\sc Fig.~8 ---} Distribution of
the luminosity-corrected color difference \dbr$_{corr}$ for star-forming
disk galaxies with and without morphological peculiarities. \dbr$_{corr}$
is defined as the residual in \dbr\ after subtracting the fitted relation
shown in Fig.~3a.} } }



\noindent\leavevmode
\noindent\makebox[0.485\textwidth]{
   \centerline{
      \psfig{file=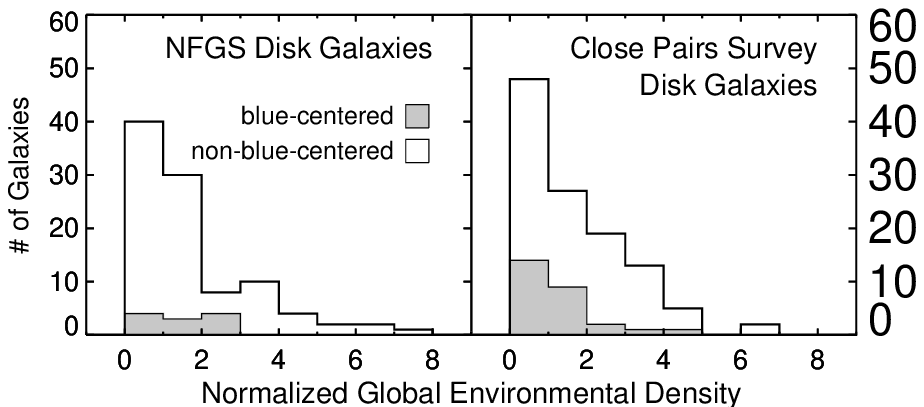,width=0.42\textwidth,clip=}
   }
}\par
\noindent\makebox[0.485\textwidth]{
   \centerline{
\parbox[t]{0.485\textwidth}{\footnotesize {\sc Fig.~9 ---} Distributions of
global environmental density for blue-centered and non-blue-centered
galaxies.  Densities are expressed in units of the mean density of galaxies
brighter than M$_{\rm B}\sim-17$ smoothed on 6.7 Mpc scales, using code
adapted from N. Grogin \citep{grogin.geller:lyalpha}.  In these units the
densities of the Virgo and Coma clusters are $\sim$4.9 and 7.4
respectively.  } } }



\noindent\leavevmode
\noindent\makebox[0.485\textwidth]{
   \centerline{
      \psfig{file=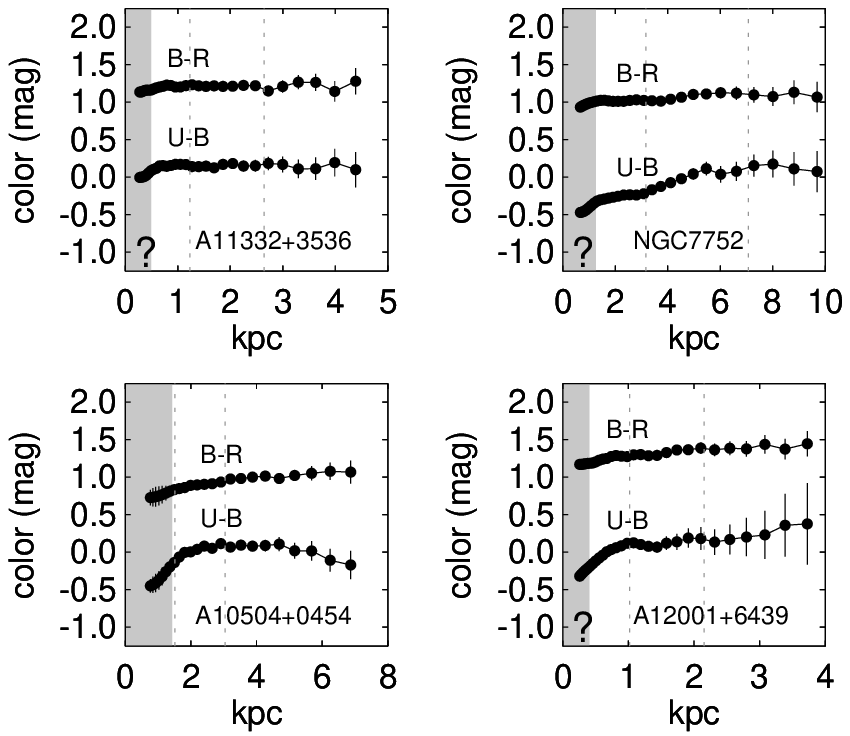,width=0.42\textwidth,clip=}
   }
}\par
\noindent\makebox[0.485\textwidth]{
   \centerline{
\parbox[t]{0.485\textwidth}{\footnotesize {\sc Fig.~10 ---} Radial color
profiles for four blue-centered galaxies that show stronger central blueing
in $U-B$ than they do in $B-R$, indicating radial gradients in starburst
properties (\S~\ref{sc:plausbulge}).  Color profiles are annotated as in
Fig.~1.} } }\\



\noindent\leavevmode
\noindent\makebox[0.485\textwidth]{
   \centerline{
      \psfig{file=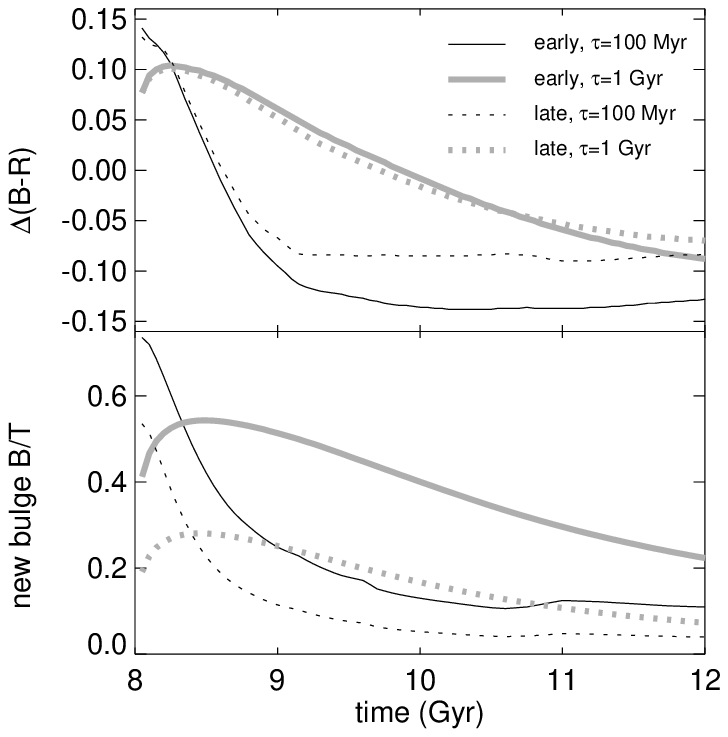,width=0.42\textwidth,clip=}
   }
}\par
\noindent\makebox[0.485\textwidth]{
   \centerline{
\parbox[t]{0.485\textwidth}{\footnotesize {\sc Fig.~11 ---} Time evolution
of \dbr\ and of the $B/T$ contribution from new bulge growth for model
blue-centered galaxies.  See \S~\ref{sc:plausgrowth}.  Solid lines indicate
early-type galaxy models, and dotted lines indicate late-type galaxy
models.  Thin black lines represent short burst timescales for the new
bulge growth ($\tau=100$Myr) while thick gray lines represent long burst
timescales ($\tau=1$Gyr). } } }\\



\noindent\leavevmode
\noindent\makebox[\textwidth]{
   \centerline{
      \psfig{file=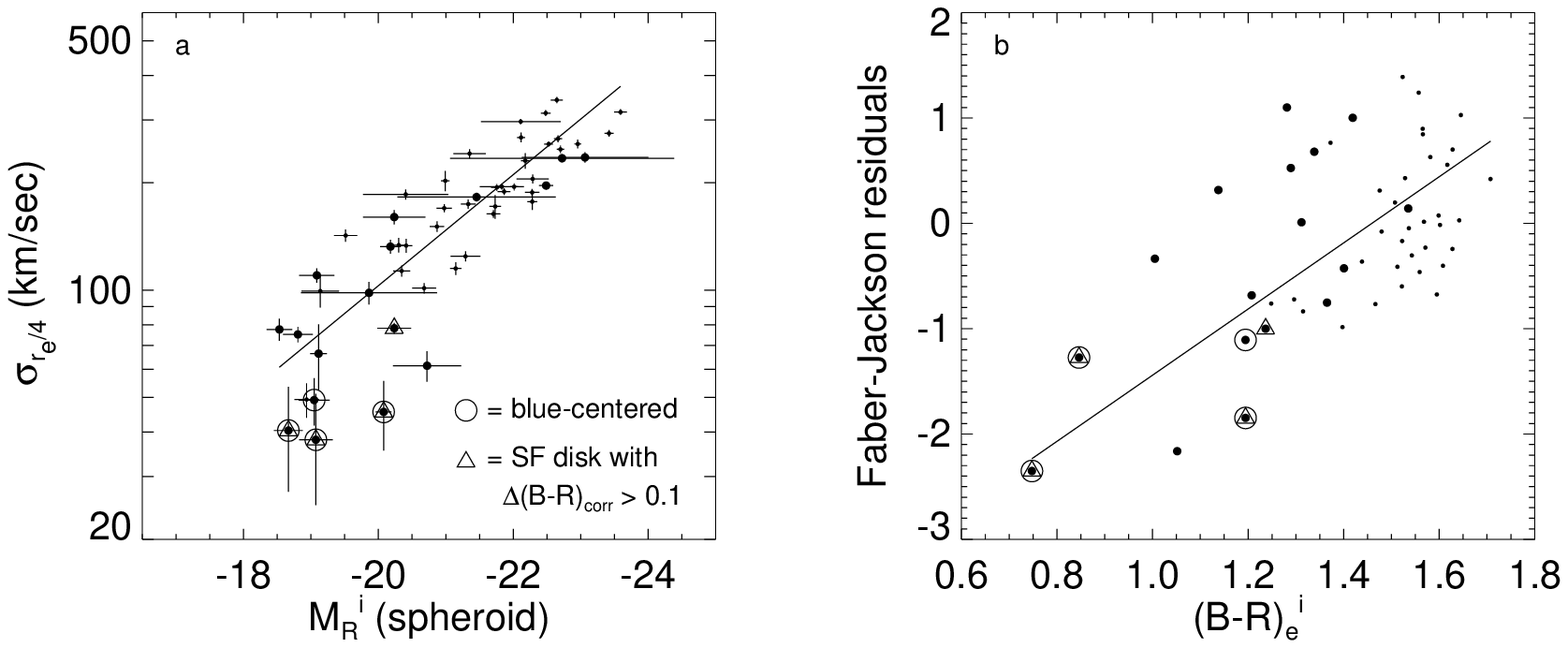,width=0.8\textwidth,clip=}
   }
}\par
\noindent\makebox[\textwidth]{
   \centerline{
\parbox[t]{\textwidth}{\footnotesize {\sc Fig.~12 ---} (\emph{a})
Faber-Jackson relation for bulges and elliptical galaxies in the NFGS.
Large dots represent bulges of star-forming disk galaxies.  Small dots
represent elliptical galaxies and non-star-forming disk galaxies, which
define a tighter relation.  Circles indicate blue-centered galaxies,
and triangles mark galaxies satisfying \dbr$_{corr}> 0.1$ (see
\S~\ref{sc:drivers}).  The latter criterion adds the
almost-blue-centered galaxy A22551+1931N to the set of galaxies likely
to be experiencing active bulge growth.  This galaxy has a companion
at 11.5 kpc listed in the CfA~2 Survey but below the detection limit
of the UZC. The line is a least-squares bisector fit excluding
blue-centered galaxies.  (\emph{b}) Faber-Jackson residuals vs.\
$B-R$ colors within $r_e$.  The line is a least-squares bisector fit
excluding blue-centered galaxies, with slope 3.1.  The slope
would be 3.5 without extinction corrections \citep[performed on star-forming disk galaxies only, using the prescriptions of][]{tully.pierce.ea:global}.
 } } }\\


\begin{thebibliography}{68}
\expandafter\ifx\csname natexlab\endcsname\relax\def\natexlab#1{#1}\fi

\bibitem[{{Abraham} {et~al.}(1996){Abraham}, {Tanvir}, {Santiago}, {Ellis},
  {Glazebrook}, \& {van den Bergh}}]{abraham.tanvir.ea:galaxy}
{Abraham}, R.~G., {Tanvir}, N.~R., {Santiago}, B.~X., {Ellis}, R.~S.,
  {Glazebrook}, K., \& {van den Bergh}, S. 1996, \mnras, 279, L47

\bibitem[{{Aguerri} {et~al.}(2001){Aguerri}, {Balcells}, \&
  {Peletier}}]{aguerri.balcells.ea:growth}
{Aguerri}, J.~A.~L., {Balcells}, M., \& {Peletier}, R.~F. 2001, \aap, 367, 428

\bibitem[{{Andredakis} {et~al.}(1995){Andredakis}, {Peletier}, \&
  {Balcells}}]{andredakis.peletier.ea:shape}
{Andredakis}, Y.~C., {Peletier}, R.~F., \& {Balcells}, M. 1995, \mnras, 275,
  874

\bibitem[{{B{\" o}ker} {et~al.}(2003){B{\" o}ker}, {Stanek}, \& {van der
  Marel}}]{b-oker.stanek.ea:searching}
{B{\" o}ker}, T., {Stanek}, R., \& {van der Marel}, R.~P. 2003, \aj, 125, 1073

\bibitem[{{Barnes}(2002)}]{barnes:formation}
{Barnes}, J.~E. 2002, \mnras, 333, 481

\bibitem[{{Barton} {et~al.}(2000){Barton}, {Geller}, \&
  {Kenyon}}]{barton.geller.ea:tidally}
{Barton}, E.~J., {Geller}, M.~J., \& {Kenyon}, S.~J. 2000, \apj, 530, 660

\bibitem[{{Barton} \& {van Zee}(2001)}]{barton.:possible}
{Barton}, E.~J. \& {van Zee}, L. 2001, \apjl, 550, L35

\bibitem[{{Barton Gillespie} {et~al.}(2003){Barton Gillespie}, {Geller}, \&
  {Kenyon}}]{barton-gillespie.geller.ea:tidally}
{Barton Gillespie}, E., {Geller}, M.~J., \& {Kenyon}, S.~J. 2003, \apj, 582,
  668

\bibitem[{{Becker} {et~al.}(2003){Becker}, {Helfand}, {White}, {Gregg}, \&
  {Laurent-Muehleisen}}]{becker.helfand.ea:first}
{Becker}, R.~H., {Helfand}, D.~J., {White}, R.~L., {Gregg}, M.~D., \&
  {Laurent-Muehleisen}, S.~A. 2003, VizieR Online Data Catalog, 8071, 0

\bibitem[{{Bruzual} \& {Charlot}(1996)}]{bruzual-a-.charlot:models}
{Bruzual A.}, G. \& {Charlot}, S. 1996, in AAS CD-ROM Series, Vol. 7,
  Astrophysics on Disc (Washington: AAS)

\bibitem[{{Carollo}(1999)}]{carollo:centers}
{Carollo}, C.~M. 1999, \apj, 523, 566

\bibitem[{{Carollo} {et~al.}(2001){Carollo}, {Stiavelli}, {de Zeeuw}, {Seigar},
  \& {Dejonghe}}]{carollo.stiavelli.ea:hubble}
{Carollo}, C.~M., {Stiavelli}, M., {de Zeeuw}, P.~T., {Seigar}, M., \&
  {Dejonghe}, H. 2001, \apj, 546, 216

\bibitem[{{Conselice}(2003)}]{conselice:relationship}
{Conselice}, C.~J. 2003, \apjs, 147, 1

\bibitem[{{Courteau} {et~al.}(1996){Courteau}, {de Jong}, \&
  {Broeils}}]{courteau..ea:evidence}
{Courteau}, S., {de Jong}, R.~S., \& {Broeils}, A.~H. 1996, \apjl, 457, L73

\bibitem[{{Cowie} {et~al.}(1996){Cowie}, {Songaila}, {Hu}, \&
  {Cohen}}]{cowie.songaila.ea:new}
{Cowie}, L.~L., {Songaila}, A., {Hu}, E.~M., \& {Cohen}, J.~G. 1996, \aj, 112,
  839

\bibitem[{{Eggen} {et~al.}(1962){Eggen}, {Lynden-Bell}, \&
  {Sandage}}]{eggen.lynden-bell.ea:evidence}
{Eggen}, O.~J., {Lynden-Bell}, D., \& {Sandage}, A.~R. 1962, \apj, 136, 748

\bibitem[{{Ellis} {et~al.}(2001){Ellis}, {Abraham}, \&
  {Dickinson}}]{ellis.abraham.ea:relative}
{Ellis}, R.~S., {Abraham}, R.~G., \& {Dickinson}, M. 2001, \apj, 551, 111

\bibitem[{{Faber} \& {Jackson}(1976)}]{faber.jackson:velocity}
{Faber}, S.~M. \& {Jackson}, R.~E. 1976, \apj, 204, 668

\bibitem[{{Falco} {et~al.}(1999){Falco}, {Kurtz}, {Geller}, {Huchra}, {Peters},
  {Berlind}, {Mink}, {Tokarz}, \& {Elwell}}]{falco.kurtz.ea:updated}
{Falco}, E.~E., {Kurtz}, M.~J., {Geller}, M.~J., {Huchra}, J.~P., {Peters}, J.,
  {Berlind}, P., {Mink}, D.~J., {Tokarz}, S.~P., \& {Elwell}, B. 1999, \pasp,
  111, 438

\bibitem[{{Friedli} \& {Benz}(1993)}]{friedli.benz:secular}
{Friedli}, D. \& {Benz}, W. 1993, \aap, 268, 65

\bibitem[{{Geller} \& {Huchra}(1989)}]{geller.huchra:mapping}
{Geller}, M.~J. \& {Huchra}, J.~P. 1989, Science, 246, 897

\bibitem[{{Gordon} {et~al.}(1997){Gordon}, {Calzetti}, \&
  {Witt}}]{gordon.calzetti.ea:dust}
{Gordon}, K.~D., {Calzetti}, D., \& {Witt}, A.~N. 1997, \apj, 487, 625

\bibitem[{{Graham}(2001)}]{graham:investigation}
{Graham}, A.~W. 2001, \aj, 121, 820

\bibitem[{{Grogin} \& {Geller}(1998)}]{grogin.geller:lyalpha}
{Grogin}, N.~A. \& {Geller}, M.~J. 1998, \apj, 505, 506

\bibitem[{{Grogin} \& {Geller}(2000)}]{grogin.geller:imaging*1}
---. 2000, \aj, 119, 32

\bibitem[{{Guzman} {et~al.}(1996){Guzman}, {Koo}, {Faber}, {Illingworth},
  {Takamiya}, {Kron}, \& {Bershady}}]{guzman.koo.ea:on}
{Guzman}, R., {Koo}, D.~C., {Faber}, S.~M., {Illingworth}, G.~D., {Takamiya},
  M., {Kron}, R.~G., \& {Bershady}, M.~A. 1996, \apjl, 460, L5

\bibitem[{{Hernquist} \& {Mihos}(1995)}]{hernquist.mihos:excitation}
{Hernquist}, L. \& {Mihos}, J.~C. 1995, \apj, 448, 41

\bibitem[{{Huchra} {et~al.}(1983){Huchra}, {Davis}, {Latham}, \&
  {Tonry}}]{huchra.davis.ea:survey}
{Huchra}, J., {Davis}, M., {Latham}, D., \& {Tonry}, J. 1983, \apjs, 52, 89

\bibitem[{{Jansen}(2000)}]{jansen:nearby}
{Jansen}, R.~A. 2000, PhD thesis, Univ.\ Groningen, The Netherlands

\bibitem[{{Jansen} {et~al.}(2000{\natexlab{a}}){Jansen}, {Fabricant}, {Franx},
  \& {Caldwell}}]{jansen.fabricant.ea:spectrophotometry}
{Jansen}, R.~A., {Fabricant}, D., {Franx}, M., \& {Caldwell}, N.
  2000{\natexlab{a}}, \apjs, 126, 331

\bibitem[{{Jansen} {et~al.}(2000{\natexlab{b}}){Jansen}, {Franx}, {Fabricant},
  \& {Caldwell}}]{jansen.franx.ea:surface}
{Jansen}, R.~A., {Franx}, M., {Fabricant}, D., \& {Caldwell}, N.
  2000{\natexlab{b}}, \apjs, 126, 271

\bibitem[{{Kannappan}(2001)}]{kannappan:kinematic}
{Kannappan}, S.~J. 2001, PhD thesis, Harvard University

\bibitem[{{Kannappan} \& {Barton}(2003)}]{kannappan.barton:tools}
{Kannappan}, S.~J. \& {Barton}, E.~J. 2003, \aj, submitted

\bibitem[{{Kannappan} \& {Fabricant}(2001)}]{kannappan.fabricant:broad}
{Kannappan}, S.~J. \& {Fabricant}, D.~G. 2001, \aj, 121, 140

\bibitem[{{Kannappan} {et~al.}(2002){Kannappan}, {Fabricant}, \&
  {Franx}}]{kannappan.fabricant.ea:physical}
{Kannappan}, S.~J., {Fabricant}, D.~G., \& {Franx}, M. 2002, \aj, 123, 2358

\bibitem[{{Kauffmann}(1996)}]{kauffmann:age}
{Kauffmann}, G. 1996, \mnras, 281, 487

\bibitem[{{Keel} {et~al.}(1985){Keel}, {Kennicutt}, {Hummel}, \& {van der
  Hulst}}]{keel.kennicutt.ea:effects}
{Keel}, W.~C., {Kennicutt}, R.~C., {Hummel}, E., \& {van der Hulst}, J.~M.
  1985, \aj, 90, 708

\bibitem[{{Kennicutt}(1998)}]{kennicutt:star}
{Kennicutt}, R.~C., J. 1998, \araa, 36, 189

\bibitem[{{Kennicutt} \& {Keel}(1984)}]{kennicutt.keel:induced}
{Kennicutt}, R.~C. \& {Keel}, W.~C. 1984, \apjl, 279, L5

\bibitem[{{Kobulnicky} \& {Zaritsky}(1999)}]{kobulnicky.zaritsky:chemical}
{Kobulnicky}, H.~A. \& {Zaritsky}, D. 1999, \apj, 511, 118

\bibitem[{{Kormendy}(1993)}]{kormendy:kinematics}
{Kormendy}, J. 1993, in IAU Symp. 153: Galactic Bulges, Vol. 153, 209

\bibitem[{{Kormendy} \& {Illingworth}(1983)}]{kormendy.illingworth:l}
{Kormendy}, J. \& {Illingworth}, G. 1983, \apj, 265, 632

\bibitem[{{Kraan-Korteweg} {et~al.}(1984){Kraan-Korteweg}, {Sandage}, \&
  {Tammann}}]{kraan-korteweg.sandage.ea:effect}
{Kraan-Korteweg}, R.~C., {Sandage}, A., \& {Tammann}, G.~A. 1984, \apj, 283, 24

\bibitem[{{Larson} \& {Tinsley}(1978)}]{larson.tinsley:star}
{Larson}, R.~B. \& {Tinsley}, B.~M. 1978, \apj, 219, 46

\bibitem[{{MacArthur} {et~al.}(2002){MacArthur}, {Courteau}, \&
  {Holtzman}}]{macarthur.courteau.ea:structure}
{MacArthur}, L.~A., {Courteau}, S., \& {Holtzman}, J.~A. 2002, in \apj,
  accepted, 8404

\bibitem[{{Maoz} {et~al.}(1995){Maoz}, {Filippenko}, {Ho}, {Rix}, {Bahcall},
  {Schneider}, \& {Macchetto}}]{maoz.filippenko.ea:detection}
{Maoz}, D., {Filippenko}, A.~V., {Ho}, L.~C., {Rix}, H., {Bahcall}, J.~N.,
  {Schneider}, D.~P., \& {Macchetto}, F.~D. 1995, \apj, 440, 91

\bibitem[{{Mihos} \& {Hernquist}(1994)}]{mihos.hernquist:triggering}
{Mihos}, J.~C. \& {Hernquist}, L. 1994, \apjl, 425, L13

\bibitem[{{Moriondo} {et~al.}(1998){Moriondo}, {Giovanardi}, \&
  {Hunt}}]{moriondo.giovanardi.ea:near-infrared}
{Moriondo}, G., {Giovanardi}, C., \& {Hunt}, L.~K. 1998, \aaps, 130, 81

\bibitem[{{Noguchi}(1996)}]{noguchi:barred}
{Noguchi}, M. 1996, \apj, 469, 605

\bibitem[{{Peletier} \& {Balcells}(1996)}]{peletier.balcells:ages}
{Peletier}, R.~F. \& {Balcells}, M. 1996, \aj, 111, 2238

\bibitem[{{Peletier} {et~al.}(1999){Peletier}, {Balcells}, {Davies},
  {Andredakis}, {Vazdekis}, {Burkert}, \&
  {Prada}}]{peletier.balcells.ea:galactic}
{Peletier}, R.~F., {Balcells}, M., {Davies}, R.~L., {Andredakis}, Y.,
  {Vazdekis}, A., {Burkert}, A., \& {Prada}, F. 1999, \mnras, 310, 703

\bibitem[{{Pfenniger} \& {Norman}(1990)}]{pfenniger.norman:dissipation}
{Pfenniger}, D. \& {Norman}, C. 1990, \apj, 363, 391

\bibitem[{{Prugniel} {et~al.}(2001){Prugniel}, {Maubon}, \&
  {Simien}}]{prugniel.maubon.ea:formation}
{Prugniel}, P., {Maubon}, G., \& {Simien}, F. 2001, \aap, 366, 68

\bibitem[{{Rix} {et~al.}(1995){Rix}, {Kennicutt}, {Braun}, \&
  {Walterbos}}]{rix.kennicutt.ea:placid}
{Rix}, H. W.~R., {Kennicutt}, R.~C., J., {Braun}, R., \& {Walterbos}, R. A.~M.
  1995, \apj, 438, 155

\bibitem[{{Scannapieco} \& {Tissera}(2003)}]{scannapieco.tissera:effects}
{Scannapieco}, C. \& {Tissera}, P.~B. 2003, \mnras, 338, 880

\bibitem[{{Schlegel} {et~al.}(1998){Schlegel}, {Finkbeiner}, \&
  {Davis}}]{schlegel.finkbeiner.ea:maps}
{Schlegel}, D.~J., {Finkbeiner}, D.~P., \& {Davis}, M. 1998, \apj, 500, 525

\bibitem[{{Schwarzkopf} \& {Dettmar}(2000)}]{schwarzkopf.dettmar:influence*1}
{Schwarzkopf}, U. \& {Dettmar}, R.-J. 2000, \aap, 361, 451

\bibitem[{{Schweizer}(1990)}]{schweizer:interactions}
{Schweizer}, F. 1990, in Dynamics and Interactions of Galaxies, ed. R. Wielen.
  Springer-Verlag (Berlin, Heidelberg), 60--71

\bibitem[{{Schweizer} \& {Seitzer}(1992)}]{schweizer.seitzer:correlations}
{Schweizer}, F. \& {Seitzer}, P. 1992, \aj, 104, 1039

\bibitem[{{Sellwood}(2000)}]{sellwood:most}
{Sellwood}, J.~A. 2000, in ASP Conf. Ser. 197: Dynamics of Galaxies: from the
  Early Universe to the Present, 3

\bibitem[{{Shioya} {et~al.}(1998){Shioya}, {Tosaki}, {Ohyama}, {Murayama},
  {Yamada}, {Ishizuki}, \& {Taniguchi}}]{shioya.tosaki.ea:molecular}
{Shioya}, Y., {Tosaki}, T., {Ohyama}, Y., {Murayama}, T., {Yamada}, T.,
  {Ishizuki}, S., \& {Taniguchi}, Y. 1998, \pasj, 50, 317

\bibitem[{{Steinmetz} \& {Navarro}(2002)}]{steinmetz.navarro:hierarchical}
{Steinmetz}, M. \& {Navarro}, J.~F. 2002, New Astronomy, 7, 155

\bibitem[{{Tissera} {et~al.}(2002){Tissera}, {Dom{\'{\i}}nguez-Tenreiro},
  {Scannapieco}, \& {S{\' a}iz}}]{tissera.domnguez-tenreiro.ea:double}
{Tissera}, P.~B., {Dom{\'{\i}}nguez-Tenreiro}, R., {Scannapieco}, C., \& {S{\'
  a}iz}, A. 2002, \mnras, 333, 327

\bibitem[{{Tully} \& {Fisher}(1977)}]{tully.fisher:new}
{Tully}, R.~B. \& {Fisher}, J.~R. 1977, \aap, 54, 661

\bibitem[{{Tully} {et~al.}(1998){Tully}, {Pierce}, {Huang}, {Saunders},
  {Verheijen}, \& {Witchalls}}]{tully.pierce.ea:global}
{Tully}, R.~B., {Pierce}, M.~J., {Huang}, J., {Saunders}, W., {Verheijen}, M.
  A.~W., \& {Witchalls}, P.~L. 1998, \aj, 115, 2264

\bibitem[{{Tully} {et~al.}(1996){Tully}, {Verheijen}, {Pierce}, {Huang}, \&
  {Wainscoat}}]{tully.verheijen.ea:ursa}
{Tully}, R.~B., {Verheijen}, M. A.~W., {Pierce}, M.~J., {Huang}, J., \&
  {Wainscoat}, R.~J. 1996, \aj, 112, 2471

\bibitem[{{Walker} {et~al.}(1996){Walker}, {Mihos}, \&
  {Hernquist}}]{walker.mihos.ea:quantifying}
{Walker}, I.~R., {Mihos}, J.~C., \& {Hernquist}, L. 1996, \apj, 460, 121

\bibitem[{{Wyse} {et~al.}(1997){Wyse}, {Gilmore}, \&
  {Franx}}]{wyse.gilmore.ea:galactic}
{Wyse}, R.~F.~G., {Gilmore}, G., \& {Franx}, M. 1997, \araa, 35, 637

\end{thebibliography}
\end{document}